\documentclass[lettersize,journal]{IEEEtran}

\usepackage{amsmath,amssymb,amsfonts}
\usepackage{algorithmic}
\usepackage{graphicx}
\usepackage{booktabs}
\usepackage{multirow}
\usepackage{textcomp}
\usepackage{xcolor}
\usepackage{tikz}
\usetikzlibrary{positioning,calc,arrows.meta}
\usepackage[hidelinks]{hyperref}
\hypersetup{
  pdftitle={Emotion as a Distribution: Joint Valence-Arousal Probability
    Learning for Speaker-Independent Multimodal Emotion Recognition},
  pdfauthor={Ting-Yi Lin, Wen-Ren Yang, and Kuanwei Chen}}

\newcommand{\valence}{\ensuremath{v}}
\newcommand{\arousal}{\ensuremath{a}}

\begin{document}

\title{Emotion as a Distribution: Joint Valence--Arousal
Probability Learning for Speaker-Independent Multimodal
Emotion Recognition}

\author{Tingyi~Lin, Wen-Ren~Yang, and~Kuanwei~Chen%
\thanks{T. Lin and W.-R. Yang are with the Department of Electrical
Engineering, National Changhua University of Education, Changhua, Taiwan
(e-mail: M1452024@mail.ncue.edu.tw; wry87c@cc.ncue.edu.tw).}%
\thanks{K. Chen is with the Department of Computer Science and Information
Engineering, National Central University, Taoyuan, Taiwan.}%
\thanks{Corresponding author: Tingyi Lin.}}

\markboth{Preprint --- submitted to \emph{Speech Communication}, 2026}%
{Lin \MakeLowercase{\textit{et al.}}: Emotion as a Distribution: Joint V--A Probability Learning for Multimodal Emotion Recognition}

\maketitle

\begin{abstract}
Human emotion is graded and frequently mixed, yet most multimodal
recognizers collapse it onto a single hard label. We argue the recognizer
should instead expose a \emph{distribution} over affective space. Our
text$+$speech system, alongside its categorical decision, emits a
$9\times9$ probability matrix over the Valence--Arousal plane, trained
with a two-dimensional Gaussian soft target under a
Kullback--Leibler/cross-entropy objective, aimed at
counseling support. Evaluation is strict:
speaker-independent 5-fold leave-one-session-out IEMOCAP with
rotating-session inner validation, headline metrics only on the held-out
session. Within one fixed encoder--fusion--head pipeline we compare
Transformer and state-space (Mamba-1/2/3) backbones at matched
depth and width, at two operating points ($T\!\approx\!550$,
$T\!\approx\!2750$). The featured dual-head system reaches
$73.0\%\!\pm\!0.3$ unweighted accuracy over three seeds (separate
rerun: $72.1\%$), exceeding the Transformer fusion baseline by
$3.0$ UA points ($95\%$ session-bootstrap CI $[1.0,4.7]$; significant
under paired $t$-test and session-level bootstrap), with no
latency or memory advantage at these lengths; swapping the $\sim$1M trainable front-end for frozen
WavLM-Large features (learnable layer weights) lifts the same
architecture to $76.6\%\!\pm\!1.3$. Pre-specified controls scope the
claims honestly: simpler valence--arousal auxiliaries reproduce the
classification lift within noise, and a dedicated regression head tracks
the continuous ratings slightly better, so the head's specific value is
the normalized affect distribution itself. That distribution recovers
the circumplex: its center of mass tracks valence and arousal
(CCC~$0.66/0.66$; predominantly between-class structure, weaker
within-class tracking), and its entropy is weakly but consistently
linked to categorical rater ambiguity, not dimensional spread.
\end{abstract}


\section{Introduction}\label{sec:intro}

\IEEEPARstart{P}{sychological} distress is increasingly common, while access to professional
counseling remains limited, motivating automated systems that can provide
timely, low-barrier emotional support and assist practitioners in tracking a
client's affective state~\cite{Kalateh2024MERSurvey}. Multimodal emotion
recognition (MER) from text and speech is a natural building block for such
systems: it is non-invasive and deployable from ordinary recordings. (In this
Phase-1 study we use only text and speech; a fuller treatment of privacy and of
physiological signals belongs to future work, and we make no privacy claim
beyond the modalities used here.)

Despite rapid progress, two issues limit how useful current MER systems are in
affective settings. \emph{First}, most systems treat emotion as a single
mutually-exclusive class. Real affective states are graded and frequently
\emph{mixed} --- an utterance can be simultaneously sad and angry --- and a hard
$\arg\max$ label compresses this graded, geometry-bearing affect into a single
class, discarding information an empathetic downstream component would otherwise
reason over. \emph{Second}, reported accuracies are
often inflated by weak evaluation: speaker-dependent splits, or model selection
on the test set, leak speaker identity and over-estimate generalization to
unseen speakers, which is the condition that actually matters in deployment.

We address both. Beyond a single class label, our recognizer emits a $9\times9$
probability matrix over the Valence--Arousal (V--A) plane --- a normalized
distribution over the discretized V--A grid. The matrix is supervised with a
two-dimensional Gaussian soft target placed at the ground-truth $(\valence,
\arousal)$ coordinate, casting emotion recognition as a label-distribution
problem whose output \emph{can} place mass on multiple regions of the affect
plane instead of being forced onto a single class. We pair this with a
deliberately strict, speaker-independent evaluation protocol --- 5-fold
leave-one-session-out (LOSO) with rotating-session inner validation --- in which
model selection happens only on validation and the headline number is computed
once on a held-out speaker session. Within one fixed encoder--fusion--head
pipeline we additionally compare Transformer and
state-space (Mamba-1/2/3~\cite{Gu2023Mamba,Dao2024SSD,Lahoti2026Mamba3}) temporal backbones
at matched depth and width, so that the major architectural confounds ---
encoder, fusion, heads, and training protocol --- are held fixed (parameter
counts differ at the block level; we report them alongside the results).

\noindent\textbf{Contributions.}
\begin{itemize}
  \item A \emph{probabilistic} V--A representation: a $9\times9$ soft-label head
        trained with a 2D-Gaussian target and a KL\,+\,cross-entropy objective
        that emits a full distribution over the V--A plane alongside the
        class decision, rather than collapsing to a single label; its
        per-utterance entropy is weakly but consistently (edge-controlled, all
        seeds) associated with human categorical ambiguity
        (Sec.~\ref{sec:method-vahead}).
  \item A \emph{rigorous} speaker-independent evaluation protocol (LOSO with
        rotating-session inner validation) that we argue should be the reporting
        standard for this task (Sec.~\ref{sec:setup-protocol}).
  \item A study of Transformer vs.\ Mamba-1/2/3 temporal backbones under one
        fixed pipeline at matched depth and width on IEMOCAP, with a direct
        characterization of the accuracy/cost trade-off (Sec.~\ref{sec:results}):
        the state-space backbones are numerically on par with or slightly above
        the Transformer in accuracy yet hold no latency or memory advantage at
        these utterance lengths, and the probabilistic V--A head adds its
        distributional output at no cost to classification accuracy.
\end{itemize}

\section{Related Work}\label{sec:related}

\subsection{Multimodal and Speech Emotion Recognition}
Early emotion recognizers were built on CNNs, LSTMs, and, more recently,
Transformer architectures, with a substantial line of work on how to
\emph{fuse} acoustic and lexical streams~\cite{Shang2024SpeechText,
Yerragondu2025MFAV, Kalateh2024MERSurvey}. Self-supervised speech representations such as
Wav2Vec2~\cite{Baevski2020Wav2Vec2} and the SER-specialized
emotion2vec~\cite{Ma2024Emotion2vec} now provide strong acoustic features, and
most competitive systems combine them with a lexical encoder through attention.
Our fusion is a standard two-stream cross-attention, which we keep fixed so that
the study isolates the temporal backbone rather than the fusion design; recent
state-space SER such as TF-Mamba~\cite{Zhao2025TFMamba}, which applies Mamba-2
across the temporal and frequency axes on IEMOCAP, and the multimodal
MaTAV~\cite{Li2025MaTAV}, a Mamba text--audio--video network for conversational
emotion recognition on IEMOCAP/MELD, are the closest model-side peers and
motivate our use of Mamba as a backbone; we differ in targeting utterance-level
text$+$speech under a strict speaker-independent protocol rather than
conversation-level recognition with video.

A recurring obstacle to comparing these systems is the \emph{evaluation
protocol}. Reported IEMOCAP accuracies range widely because some works use
random utterance-level $k$-fold splits (which leak speaker identity and reach
$\sim$$78$--$80\%$ WA), while others use the harder speaker-independent
leave-one-session-out (LOSO) setting ($\sim$$70$--$75\%$ UA). The two are not
comparable. We adopt the strict speaker-independent LOSO protocol throughout
(Sec.~\ref{sec:setup-protocol}) and, when tabulating prior numbers, record the
protocol, modalities, and metric so that cross-protocol figures are never
silently compared. Table~\ref{tab:sota} compiles representative recent IEMOCAP
results together with their evaluation protocol and modalities. Among the entries we could \emph{verify} under a comparable
speaker-independent protocol, our \texttt{mamba\_dual\_head} is competitive:
$73.0$ UA (three-seed mean) against emotion2vec's $71.79$ WA with an
emotion-specialized front-end, and our own frozen WavLM-Large~\cite{Chen2022WavLM} arm reaches
$76.6\pm1.3$ UA under the same strict LOSO protocol
(Sec.~\ref{sec:results-controls}) against TF-Mamba's $75.7$ under the easier
5-fold CV; the higher figures in Table~\ref{tab:sota} use 5-fold
cross-validation or leave-one-speaker-out partitions that are not directly
comparable, and we rest no comparison on rows whose partition leaves
speaker independence unstated (DFSD-EMO, for instance, specifies only
``5-fold cross-validation'', so speaker independence cannot be established
from the source). We
further use \emph{standard} frozen encoders rather than emotion-specialized
self-supervised front-ends such as emotion2vec, so absolute accuracy is not the
contribution --- the representation and the protocol are.

\begin{table*}[t]
\centering
\caption{Representative recent IEMOCAP (4-class) results under \emph{different}
evaluation protocols. Protocol tags mark the partition family --- session
folds (LOSO; emotion2vec and ours) hold entire speaker pairs out, generic
``5-fold CV'' entries do not always specify speaker independence, and
leave-one-speaker-out (LOSpkO; SDMMKD) uses 10 finer folds --- and each
paper's protocol wording is reproduced \emph{verbatim} in the
$\dagger$-numbered table footnotes, so partition differences are visible
rather than flattened. Params (M) as reported by each source:
$a{+}b$\,=\,frozen upstream $+$ trainable probe; ``fr.''\,=\,frozen
front-end additional to the stated trainable count; n/r\,=\,not reported.
Separate WA and UA columns (--- where the source
does not report the metric; EmoBridge additionally reports wF1 $73.8$).
A\,=\,audio, T\,=\,text. Scores are taken verbatim from each source paper; the
in-table note states the merge, seed, and encoder-regime annotations. Table
body generated from the committed statistics.}
\label{tab:sota}
\begin{tabular}{@{}llp{0.185\linewidth}lccc@{}}
\toprule
Method & Mod. & Front-end & Protocol & Params (M) & WA (\%) & UA (\%) \\
\midrule
emotion2vec~\cite{Ma2024Emotion2vec}$^\star$ & A & emotion2vec (frozen, emotion SSL) & LOSO (5-fold)$^{\dagger1}$ & $93.8{+}0.2$ & 71.79 & --- \\
TF-Mamba~\cite{Zhao2025TFMamba}$^\star$ & A & WavLM-large & 5-fold CV$^{\dagger2}$ & $20.8$ ($+316$ fr.) & 75.3 & 75.7 \\
EmoBridge~\cite{Sun2025EmoBridge}$^\star$ & A+T & HuBERT (frozen) $+$ LLM (LoRA) & 5-fold CV (spk.-indep.)$^{\dagger3}$ & $\sim$$7{,}000$ (LLM) & 76.4 & 76.8 \\
SDMMKD~\cite{Zhu2025EnhancingSE}$^\star$ & A+T & HuBERT (LoRA-FT) $+$ SKEP teacher & LOSpkO (10-fold)$^{\dagger4}$ & n/r & 79.74 & 80.68 \\
DFSD-EMO~\cite{Yu2025DFSD}$^\star$ & A & 768-d pre-trained features & 5-fold CV$^{\dagger5}$ & n/r & 72.36 & 72.35 \\
\midrule
\textbf{Ours} (\texttt{mamba\_dual\_head})$^\star$ & A+T & frozen BERT $+$ raw-conv ($\sim$1M, FT) & LOSO, rotating val.\ (Sec.~\ref{sec:setup-protocol}) & $16.9$ & $71.4$ & $73.0$ \\
\textbf{Ours} (SSL arm)$^\star$ & A+T & frozen BERT $+$ frozen WavLM-Large & LOSO, rotating val.\ (Sec.~\ref{sec:results-controls}) & $16.0$ ($+316$ fr.) & $75.3$ & $76.6$ \\
\midrule
\multicolumn{7}{@{}p{0.97\linewidth}@{}}{\footnotesize Protocol wording \emph{verbatim} from each source: $^{\dagger1}$``leave-one-session-out 5-fold CV'' (Sec.~4.3); $^{\dagger2}$``five-fold cross-validation exclusively for the IEMOCAP dataset'' (Sec.~III-A2); $^{\dagger3}$``speaker-independent five-fold cross-validation'' (Sec.~IV-A); $^{\dagger4}$``standard 10-fold cross-validation (CV) with the leave-one-speaker-out scheme'' (Sec.~III-C); $^{\dagger5}$``5-fold cross-validation'' (Sec.~IV-B; 3-s clips).} \\
\multicolumn{7}{@{}p{0.97\linewidth}@{}}{\footnotesize $\star$: same 4-class \emph{hap}$+$\emph{exc} merge as ours (5{,}531 utterances; DFSD-EMO segments them into 3\,s clips). Comparison scores are the single runs reported by the source papers (no seed spread published); ours is a 3-seed mean. The comparison systems fine-tune, adapt, or distill large emotion-specialized or general SSL encoders (emotion2vec, WavLM-large, HuBERT), whereas our system trains $16.9$M parameters over frozen general-purpose encoders --- the rows contextualize the operating point; absolute accuracy is not the contribution.} \\
\bottomrule
\end{tabular}

\end{table*}

\subsection{Dimensional vs.\ Categorical Affect, and Soft Labels}
The Valence--Arousal circumplex~\cite{Russell1980Circumplex} represents affect as
continuous coordinates rather than discrete categories, which is more expressive
for graded and mixed emotion. A complementary line of work softens the
supervision itself. Label Distribution Learning (LDL)~\cite{Geng2016LDL} replaces
the one-hot target with a full distribution over labels, and soft-target training
has been shown to help on ambiguous emotional
utterances~\cite{Ando2018SoftTarget}. In SER specifically, a rater-distribution
lineage models the \emph{perception uncertainty} carried by the annotators
themselves: Han~\emph{et al.}~\cite{Han2017HardToSoft} train against soft labels
derived from inter-rater disagreement, Chou and
Lee~\cite{Chou2019EveryRating} learn jointly from every individual rater's vote,
and Wu~\emph{et al.}~\cite{Wu2022DirichletUncertainty} place utterance-specific
Dirichlet priors over the categorical label distribution; treating annotator
disagreement as signal rather than noise is likewise the perspectivist position
in subjective-annotation research~\cite{Davani2022Disagreements}. Closest to us
are two recent efforts: Le~\emph{et al.}~\cite{Le2023UncertaintyLDL} build label
distributions from valence--arousal-space \emph{neighborhoods} to relabel facial
expression \emph{categories}, and Prabhu~\emph{et al.}~\cite{Prabhu2022tDist,Prabhu2023LabelUncertainty}
--- the latter published in \emph{IEEE Transactions on Affective Computing} ---
model annotator-disagreement uncertainty in SER with a Bayesian network that
predicts a \emph{per-attribute} (valence or arousal) distribution. Beyond speech,
Yang and Chen~\cite{Yang2011MusicVADistribution} established the prediction of a
\emph{distribution over the continuous V--A plane} for music emotion, which we
regard as the closest output-space precedent in any audio domain.

Our $9\times9$ head differs from these speech systems in \emph{what it
predicts}. Prior soft-label / LDL emotion work in speech either (i) softens
\emph{categorical} targets over a
discrete emotion set~\cite{Ando2018SoftTarget,Han2017HardToSoft,Chou2019EveryRating,Wu2022DirichletUncertainty,Le2023UncertaintyLDL}, or
(ii) models \emph{per-attribute} uncertainty as two independent one-dimensional
regressions with predicted variance~\cite{Prabhu2023LabelUncertainty}. In
contrast, we predict an \emph{explicit joint two-dimensional probability mass
function over a discretized Valence$\times$Arousal grid}: a single softmax head
whose output is, by construction, a normalized affect distribution, trained with
a Gaussian soft target whose width $\sigma$ encodes annotation spread directly on
the grid --- to our knowledge the first \emph{speech$+$text} system whose output
space is the joint V--A plane itself. This makes the representation \emph{(a) joint
in form} --- a single distribution
over the Valence$\times$Arousal plane rather than two separate per-attribute
predictions; \emph{(b) form-free} in the affect plane --- the discrete joint
pmf assumes no parametric output family (the Gaussian enters only as the
shape of the \emph{training target}, Sec.~\ref{sec:method-vahead}) and
\emph{can} place mass on multiple, possibly non-adjacent regions, a representational capacity that neither a single class
label nor two unimodal marginals possess (whether the \emph{learned}
distributions exercise this capacity is an empirical question we quantify in
Sec.~\ref{sec:results-analysis}); and \emph{(c) designed to be consumable} as a
soft prompt by a downstream language model (our Phase-2 vision --- a design
intention, not yet a demonstrated result), without first
collapsing it to a class. Empirically (Sec.~\ref{sec:results-analysis}), this
distribution does more than soften the target: in aggregate it recovers the
affective circumplex and its center of mass tracks continuous valence and
arousal, evidence that the head learns class-level affect \emph{geometry}, which
the categorical-soft-label and per-attribute lines above do not examine.
A complementary precedent in speech is Zhou~\emph{et al.}~\cite{Zhou2023AVfromCategorical},
who show that a continuous arousal--valence representation can be learned
\emph{from categorical emotion labels alone}. We do not reproduce that setting
--- our soft target is constructed from the rater-mean $(v^\ast,a^\ast)$
coordinates, so the dimensional labels \emph{do} supervise the head, through the
target's location rather than through a point-regression loss --- but the head
likewise emits its continuous estimate without any explicit regression
objective, and yields a full joint distribution over the grid rather than a
single learned embedding.
Le~\emph{et al.}\ is the nearest speech-adjacent neighbor, but they use
V--A neighborhoods to relabel facial-expression \emph{categories}, whereas in our
model the \emph{output space itself is the V--A grid} for speech and text.

\subsection{State-Space Models}
Mamba~\cite{Gu2023Mamba} introduced a selective state-space model (SSM) with
input-dependent dynamics and a hardware-aware parallel scan, achieving linear
time complexity in sequence length. Its irregular, data-dependent recurrence is
less GPU-friendly than dense attention; Mamba-2 / Structured State-Space Duality
(SSD)~\cite{Dao2024SSD} reformulates the SSM so that it can be computed with
matrix multiplications, improving hardware utilization while retaining the
linear-recurrence view. Mamba-3~\cite{Lahoti2026Mamba3} further refines the
discretization (trapezoidal rule), replaces the short causal convolution with
rotary position embeddings, and adds an optional multi-input/multi-output (MIMO)
formulation, targeting long-context language modeling and decode-time
efficiency. Bidirectional variants~\cite{Zhu2024VisionMamba} and
multimodal applications~\cite{Qiao2024VLMamba} have since appeared, including
emotion-specific uses such as continuous valence--arousal regression~\cite{Liang2025MambaVA}
(a point estimate, in contrast to our distributional grid).
We use these models purely as \emph{temporal backbones}; we make no claim that
they overcome an attention ``complexity wall'' at the utterance lengths studied
here, and we report their cost explicitly in Sec.~\ref{sec:results}.

\section{Method}\label{sec:method}

\subsection{Overview}\label{sec:method-overview}
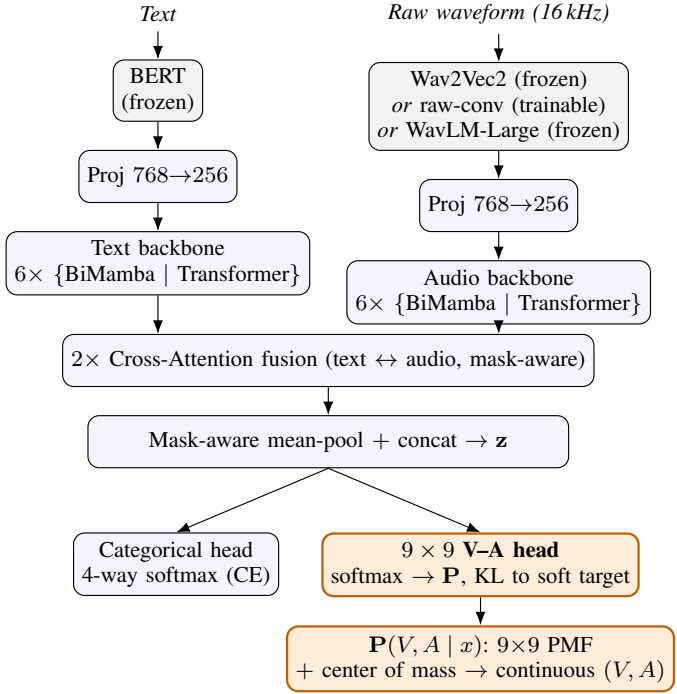
\begin{figure}[t]
  \centering
  \resizebox{\linewidth}{!}{%
  \begin{tikzpicture}[
    font=\footnotesize,
    >=Latex,
    box/.style={draw, rounded corners, align=center, minimum height=6.5mm,
                inner sep=3pt, fill=blue!4},
    enc/.style={box, fill=gray!10},
    head/.style={box, fill=orange!14, draw=orange!75!black, thick},
    io/.style={align=center, font=\footnotesize\itshape},
  ]
    \node[io] (t0) {Text};
    \node[io, right=24mm of t0] (a0) {Raw waveform (16\,kHz)};
    \node[enc, below=3.5mm of t0] (te) {BERT\\(frozen)};
    \node[enc, below=3.5mm of a0] (ae) {Wav2Vec2 (frozen)\\\emph{or} raw-conv (trainable)\\\emph{or} WavLM-Large (frozen)};
    \node[box, below=3.5mm of te] (tp) {Proj $768{\to}256$};
    \node[box, below=3.5mm of ae] (ap) {Proj $768{\to}256$};
    \node[box, below=3.5mm of tp] (tb) {Text backbone\\$6\times$ \{BiMamba $\mid$ Transformer\}};
    \node[box, below=3.5mm of ap] (ab) {Audio backbone\\$6\times$ \{BiMamba $\mid$ Transformer\}};
    \coordinate (mid) at ($(tb)!0.5!(ab)$);
    \node[box, below=7mm of mid, minimum width=60mm] (ca)
        {$2\times$ Cross-Attention fusion (text $\leftrightarrow$ audio, mask-aware)};
    \node[box, below=3.5mm of ca, minimum width=60mm] (pool)
        {Mask-aware mean-pool $+$ concat $\rightarrow \mathbf{z}$};
    \node[box, below=8mm of pool, xshift=-19mm] (h1) {Categorical head\\4-way softmax (CE)};
    \node[head, below=8mm of pool, xshift=19mm] (h2) {$9\times9$ \textbf{V--A head}\\softmax $\to \mathbf{P}$, KL to soft target};
    \node[head, below=3.5mm of h2] (out)
        {$\mathbf{P}(V,A\mid x)$: $9{\times}9$ PMF\\$+$ center of mass $\to$ continuous $(V,A)$};
    \draw[->] (t0)--(te); \draw[->] (a0)--(ae);
    \draw[->] (te)--(tp); \draw[->] (ae)--(ap);
    \draw[->] (tp)--(tb); \draw[->] (ap)--(ab);
    \draw[->] (tb.south) -- (tb.south |- ca.north);
    \draw[->] (ab.south) -- (ab.south |- ca.north);
    \draw[->] (ca)--(pool);
    \draw[->] (pool.south) -- (h1.north);
    \draw[->] (pool.south) -- (h2.north);
    \draw[->] (h2)--(out);
  \end{tikzpicture}}
  \caption{Echo-Mind Phase-1 architecture. A frozen text encoder and a
  swappable speech front-end --- frozen Wav2Vec2, fine-tuned raw-convolution,
  or frozen WavLM-Large (Sec.~\ref{sec:results-controls}) --- feed a
  shared-width projection, an interchangeable temporal backbone, bidirectional
  cross-attention fusion, and one or two output heads. The $9\times9$
  Valence--Arousal head (Sec.~\ref{sec:method-vahead}) is the paper's headline
  contribution; the categorical head is retained for comparability with prior
  work.}
  \label{fig:arch}
\end{figure}

Given an utterance, the text stream is encoded by a frozen
BERT~\cite{Devlin2019BERT} and the speech stream by an interchangeable acoustic
front-end (Sec.~\ref{sec:method-frontend}). Each modality is projected from its encoder
width ($768$) to a shared width $d=256$ by a linear layer followed by layer
normalization, GELU, and dropout, processed by a temporal
backbone (Sec.~\ref{sec:method-backbone}), fused by two stacked cross-attention
blocks (Sec.~\ref{sec:method-fusion}), and decoded by one or more heads
(Sec.~\ref{sec:method-vahead}). The pipeline is deliberately held \emph{fixed}
across all experiments in Sec.~\ref{sec:results}: only the component named by a
given ablation (acoustic front-end, temporal backbone, or output head) is varied.
Two optimization asymmetries qualify this isolation and should be borne in mind
when reading the backbone comparison: the per-category learning-rate multipliers
of Sec.~\ref{sec:setup} apply a $\times3$ rate to state-space parameters only,
and the long-sequence setting uses per-model batch sizes to fit memory
(Sec.~\ref{sec:setup}); measured differences are therefore attributable to the
varied component \emph{under this shared training recipe}, not under a per-arm
tuned optimum. The temporal backbone and the fusion block play distinct,
complementary roles: the backbone (Transformer self-attention or BiMamba) mixes
information \emph{within} each modality along time, whereas the cross-attention
blocks exchange information \emph{between} the two modalities. This division
matches each operation to its setting: intra-modal mixing runs along a single,
possibly long time axis --- the regime state-space models are designed for ---
whereas inter-modal fusion aligns the long audio stream with a short text
stream, where content-based attention is the natural primitive and, because the
text side is short, is already linear in the audio length
(Sec.~\ref{sec:method-fusion}). A state-space
backbone therefore substitutes for intra-modal self-attention, not for the
inter-modal cross-attention, which we keep fixed across all backbones.
Fig.~\ref{fig:arch} gives the overall architecture.

\subsection{Acoustic Front-End}\label{sec:method-frontend}
Speech is resampled to $16$\,kHz and amplitude-normalized (peak normalization by
default; zero-mean/unit-variance and identity are configurable alternatives). We
deliberately keep the front-end \emph{learned} rather than hand-engineered, and
expose two interchangeable encoders so that the temporal backbone can be studied
at two operating points along the time axis.

\paragraph*{SSL encoder (default, $T\!\approx\!550$)}
The default encoder is a frozen Wav2Vec2~\cite{Baevski2020Wav2Vec2} whose
convolutional feature extractor has a total stride of $320$ samples, giving a
frame rate of $50$\,fps; an $11$\,s utterance therefore yields $T\!\approx\!550$
frames of dimension $768$. Inputs pass through the HuggingFace
\texttt{Wav2Vec2FeatureExtractor} so that the encoder sees the zero-mean,
unit-variance signal it was pre-trained on. One subtlety matters for
reproducibility: this checkpoint's convolutional front-end uses \emph{temporal}
group normalization (\texttt{feat\_extract\_norm="group"}), so with dynamic
batch-max padding an utterance's features would depend on the longest utterance
that happens to share its batch. We therefore pad every batch to the fixed
$11$\,s cap, which makes the features per-utterance deterministic and
independent of batch composition.

\paragraph*{Trainable raw-convolution encoder (long context, $T\!\approx\!2750$)}
To give the temporal backbone a longer, higher-resolution sequence, we
optionally replace Wav2Vec2 with a lightweight trainable stack of three strided
1-D convolutions (channels $64\!\to\!256\!\to\!512$, kernels $8/4/4$, strides
$4/4/4$, each followed by a \emph{masked} GroupNorm and GELU, with dropout
$0.1$ between layers), projected to $768$. The GroupNorm statistics are
computed over valid frames only and padded positions are re-zeroed after every
stage; together with the per-layer mask of Eq.~\eqref{eq:conv-len} this makes
the encoder's output for an utterance independent of how much padding it
receives in a batch (plain temporal GroupNorm would mix padding into the
normalization statistics). The total
\emph{frame stride} is the product of the per-layer strides,
$s_{\mathrm{frame}}=4\cdot4\cdot4=64$, i.e.\ $250$\,fps, so the same $11$\,s
utterance now produces $T\!\approx\!2750$ frames --- a $5\times$ longer sequence
that is the regime in which state-space backbones are expected to matter. This
encoder is pre-trained on LibriSpeech~\cite{Panayotov2015LibriSpeech} by
self-supervised mel reconstruction and
then fine-tuned (or frozen) for emotion (Sec.~\ref{sec:setup-impl}).

\paragraph*{Frame-level masking}
Because utterances are batched with right padding, every model must convert the
sample-level validity mask into a frame-level mask that exactly matches the
encoder's down-sampling, otherwise padded samples leak into the attention and
the pooling. We use the convention \texttt{True}=valid throughout. For a 1-D
convolution with kernel $k$, stride $s$, padding $p$, and dilation $d$, the
number of valid output frames produced from $L_{\mathrm{in}}$ valid input
samples follows the exact PyTorch output-length formula
\begin{equation}
  L_{\mathrm{out}} = \left\lfloor
    \frac{L_{\mathrm{in}} + 2p - d\,(k-1) - 1}{s} \right\rfloor + 1 ,
  \label{eq:conv-len}
\end{equation}
which counts how many stride-$s$ steps an effective window of width $d(k-1)+1$
takes across the padded length $L_{\mathrm{in}}+2p$. We apply it layer by layer
(for the raw-conv encoder, $\{(k,s,p)\}=\{(8,4,4),(4,4,2),(4,4,2)\}$ with dilation
$d{=}1$ throughout, so this width collapses to $k$ and the numerator to
$L_{\mathrm{in}}+2p-k$) to obtain the per-utterance frame count; the frame mask
sets the first $L_{\mathrm{out}}$ positions valid and the remainder padded. The Wav2Vec2 path uses the encoder's native
\texttt{\_get\_feat\_extract\_output\_lengths}. Both encoders expose the same
\texttt{compute\_frame\_mask} interface, so swapping the front-end requires no
change in any downstream model.

\paragraph*{Why not hand-crafted spectral/prosodic inputs?}
A classical SER front-end would feed MFCCs, log-Mel spectrograms, or explicit
prosody (F0, energy, jitter) to the backbone. We avoid this for two reasons.
\emph{First}, large self-supervised speech models already encode spectral and
prosodic structure in their hidden states, and consistently outperform
hand-crafted inputs on SER; using them as the front-end removes a confound and
keeps the comparison about the \emph{backbone}, not about feature engineering.
\emph{Second}, the trainable raw-convolution encoder above lets the network learn
its own time-frequency filters end-to-end at the resolution the backbone needs,
which a fixed spectrogram cannot. We do, however, retain prosody in one
controlled place: a $12$-dimensional set of \emph{microphone-invariant} prosody
descriptors (voiced ratio, pitch mean/standard-deviation/slope/jitter, energy
contour statistics, speech rate, pause ratio, duration) can be late-fused as an
auxiliary vector. These are gain-invariant by construction --- intended for
cross-corpus and counseling-room robustness where channel and microphone differ
--- but they are \emph{disabled by default} so that the backbone and head
ablations are not confounded by an extra feature stream; we describe them
here as an implemented optional component only, and no prosody-enabled
configuration is evaluated in this paper.

\subsection{Temporal Backbone}\label{sec:method-backbone}
A selective state-space model maps an input signal $x(t)$ to an output $y(t)$
through a latent state $h(t)$ governed by the continuous-time dynamics
\begin{equation}
  h'(t) = \mathbf{A}\,h(t) + \mathbf{B}\,x(t), \qquad
  y(t)  = \mathbf{C}\,h(t).
  \label{eq:ssm-cont}
\end{equation}
To run on discrete frames, Eq.~\eqref{eq:ssm-cont} is discretized with a
zero-order hold (ZOH): the input is assumed constant, $x(\tau)\!=\!x_k$, over a
step of width $\Delta_k$. Integrating the linear ODE over $[\,t_k,t_k+\Delta_k]$
gives the closed-form solution $h(t_k+\Delta_k)=e^{\Delta_k\mathbf{A}}h(t_k)
+\big(\int_0^{\Delta_k} e^{\tau\mathbf{A}}\,d\tau\big)\mathbf{B}\,x_k$, so that
\begin{equation}
  \bar{\mathbf{A}}_k = \exp(\Delta_k\mathbf{A}), \quad
  \bar{\mathbf{B}}_k = (\Delta_k\mathbf{A})^{-1}\!\big(\exp(\Delta_k\mathbf{A})-\mathbf{I}\big)\,\Delta_k\mathbf{B},
  \label{eq:zoh}
\end{equation}
The $\Delta_k\mathbf{A}$ form of $\bar{\mathbf{B}}_k$ follows~\cite{Gu2023Mamba} and is
the closed form of the integral above,
$\big(\int_0^{\Delta_k}\!e^{\tau\mathbf{A}}\,d\tau\big)\mathbf{B}
=\mathbf{A}^{-1}\!\big(e^{\Delta_k\mathbf{A}}-\mathbf{I}\big)\mathbf{B}$. This yields the
linear recurrence and read-out
\begin{equation}
  h_k = \bar{\mathbf{A}}_k\,h_{k-1} + \bar{\mathbf{B}}_k\,x_k, \qquad
  y_k = \mathbf{C}_k\,h_k .
  \label{eq:ssm-disc}
\end{equation}
\emph{Selectivity} is what makes this a Mamba rather than a classical linear SSM:
the parameters $\mathbf{B}_k$, $\mathbf{C}_k$, and the step size $\Delta_k$ are
\emph{input-dependent} functions of $x_k$, so the model can choose, per token,
how much of the new input to write and how fast to forget --- a content-based
gating that a time-invariant SSM lacks~\cite{Gu2023Mamba}. The recurrence in
Eq.~\eqref{eq:ssm-disc} is evaluated by a hardware-aware parallel scan, giving
$\mathcal{O}(T)$ time and memory in the sequence length. Mamba-2 / Structured
State-Space Duality~\cite{Dao2024SSD} rewrites the same recurrence as a
structured matrix multiply, trading the per-token scan for chunked matmuls that
use tensor cores; we treat v1 and v2 as two realizations of the identical block
and compare them empirically (Sec.~\ref{sec:results-cost}).

\paragraph*{Bidirectional Mamba block}
We instantiate the backbone as a bidirectional Mamba (BiMamba) block: two
\emph{separate} (not weight-shared) Mamba scans process the sequence forward and
backward, and their outputs are merged by a linear map
$\mathbf{W}_M\!:\!\mathbb{R}^{2d}\!\to\!\mathbb{R}^{d}$ with a residual
connection~\cite{Zhu2024VisionMamba}. Writing $\mathrm{Mamba}_{\rightarrow}$ and
$\mathrm{Mamba}_{\leftarrow}$ for the two scans and $R_b$ for the padding-aware
valid-prefix reversal of Eq.~\eqref{eq:revop} below, the block computes
\begin{equation}
  \begin{split}
    \mathrm{BiMamba}(\mathbf{X}) = \mathbf{X} + \mathbf{W}_M\big[\,
      &\mathrm{Mamba}_{\rightarrow}(\mathrm{LN}\,\mathbf{X}) \\
      &\;\big\|\;
      R_b\!\big(\mathrm{Mamba}_{\leftarrow}(R_b(\mathrm{LN}\,\mathbf{X}))\big)\,\big],
  \end{split}
  \label{eq:bimamba}
\end{equation}
where $[\,\cdot\,\|\,\cdot\,]$ concatenates along the feature axis (a dropout on the
merged output is omitted for brevity). The same block structure is also realized with
the Mamba-2/SSD formulation~\cite{Dao2024SSD} (with its own default state size), and
compared against a standard Transformer encoder under matched depth and width.

\paragraph*{Padding-aware backward scan}
With right-padded batches, a naive backward scan $\mathrm{flip}\circ
\mathrm{Mamba}\circ\mathrm{flip}$ is subtly wrong: flipping the padded sequence
places the padding \emph{before} the valid frames, so the backward recurrence
accumulates state over padding before it ever reads real input, contaminating
every valid position (the attention baseline, by contrast, masks padded keys
exactly). We therefore reverse only the valid prefix with the index map $R_b$,
\begin{equation}
  R_b(t) = \begin{cases} T_b-1-t, & t < T_b,\\[2pt] t, & t \ge T_b,\end{cases}
  \qquad R_b\circ R_b = \mathrm{id},
  \label{eq:revop}
\end{equation}
for a sample with $T_b$ valid frames out of $T$; because $R_b$ is an involution,
applying it once reverses the valid region and applying it again restores the
original time order after the scan, so each backward output position aligns with
the same time step of the forward branch before the two are concatenated in
Eq.~\eqref{eq:bimamba}. The
forward scan needs no correction (it is causal, and right padding never
precedes valid frames); padded positions are excluded from all subsequent
attention and pooling by the validity mask. A padding-invariance regression
test (identical utterance alone vs.\ inside a longer-padded batch) guards this
property for all three Mamba generations.

\subsection{Cross-Attention Fusion}\label{sec:method-fusion}
Let $\mathbf{H}^{a}\!\in\!\mathbb{R}^{T_a\times d}$ and
$\mathbf{H}^{t}\!\in\!\mathbb{R}^{T_t\times d}$ be the audio and text backbone
outputs. Each cross-attention block lets one modality query the other and vice
versa. We write $\mathrm{MHA}(Q,K,V;m)$ for standard multi-head attention~\cite{Vaswani2017Attention} with
$h{=}8$ heads of width $d_h=d/h$: each head applies scaled dot-product attention
over the valid keys,
\begin{equation}
  \mathrm{Attn}(Q,K,V;m) = \mathrm{softmax}\!\Big(\tfrac{Q K^{\!\top}}{\sqrt{d_h}}
    + \mathbf{M}(m)\Big)\,V,
  \label{eq:attn}
\end{equation}
where the additive mask sets $\mathbf{M}(m)_{:,j}=-\infty$ at padded keys
($m_j$ invalid) and $0$ otherwise; $\mathrm{MHA}$ projects $Q,K,V$ per head,
concatenates the $h$ head outputs, and applies an output projection. The
text-to-audio and audio-to-text updates are
\begin{align}
  \tilde{\mathbf{H}}^{t} &= \mathrm{FFN}\big(\mathrm{LN}\big(\mathbf{H}^{t}
     + \mathrm{MHA}(\mathbf{H}^{t},\mathbf{H}^{a},\mathbf{H}^{a};\,m^{a})\big)\big), \\
  \tilde{\mathbf{H}}^{a} &= \mathrm{FFN}\big(\mathrm{LN}\big(\mathbf{H}^{a}
     + \mathrm{MHA}(\mathbf{H}^{a},\tilde{\mathbf{H}}^{t},\tilde{\mathbf{H}}^{t};\,m^{t})\big)\big),
  \label{eq:cross-attn}
\end{align}
where $\mathrm{FFN}(\cdot)$ denotes a position-wise feed-forward sublayer with
its own residual connection and layer norm. The two updates are applied
\emph{sequentially}: the text stream is updated first, and the audio stream
attends to the already-updated text $\tilde{\mathbf{H}}^{t}$. Two such blocks
are stacked. Each cross-attention block scales as $\mathcal{O}(T_a T_t)$ in the
two sequence lengths (the projection and feed-forward terms add a further
$\mathcal{O}((T_a{+}T_t)\,d^2)$, likewise linear in each length); because the
text stream is capped at $T_t\le 64$ tokens, the bounded factor collapses this to
\begin{equation}
  T_t \le 64 \;\Longrightarrow\; \mathcal{O}(T_a T_t) = \mathcal{O}(T_a),
  \label{eq:cross-cost}
\end{equation}
i.e.\ the fusion is \emph{linear} in the audio length $T_a$; the only potentially
quadratic term --- self-attention over the long audio sequence --- is precisely
what the state-space backbone replaces. Each
fused stream is then reduced to a single
vector by a mask-aware mean pool over its valid positions,
\begin{equation}
  \overline{\tilde{\mathbf{H}}}
  = \frac{\sum_{t} m_t\,\tilde{\mathbf{H}}_t}{\sum_{t} m_t},
  \label{eq:mean-pool}
\end{equation}
where $m_t\!\in\!\{0,1\}$ is the (\texttt{True}=valid) mask of that stream, so
padded positions contribute neither to the numerator nor the denominator. The
fused utterance representation is the concatenation
$\mathbf{z}=[\,\overline{\tilde{\mathbf{H}}^{t}}\,\|\,\overline{\tilde{\mathbf{H}}^{a}}\,]$,
which feeds the head(s).

\paragraph*{Mask convention}
Internally every validity mask uses \texttt{True}=valid, including the
frame mask of Eq.~\eqref{eq:conv-len} and the BERT token mask. PyTorch attention,
however, expects \texttt{key\_padding\_mask} with \texttt{True}=\emph{ignore}.
Every model therefore inverts the mask, $m = \neg(\text{valid})$, immediately
before the attention call, so that a query never attends to a padded key. The
mean pooling that forms $\mathbf{z}$ likewise sums only over valid positions. We
state this explicitly because a sign error here silently lets padding leak into
both the attention and the pooled representation.

\subsection{Valence--Arousal Probability Head}\label{sec:method-vahead}
In the dual-head configuration the model keeps the standard categorical 4-way
head --- which yields the class decision and all reported UA/WA/F1 --- and adds
the V--A head described here. Both read the shared fused representation
$\mathbf{z}$ and are trained jointly (Eq.~\eqref{eq:va-loss}), so in this paper
the $9\times9$ matrix is an \emph{auxiliary} distributional output (analyzed in
Sec.~\ref{sec:results-analysis}) rather than the source of the categorical
prediction; using it directly as the decision head is left to future work
(Sec.~\ref{sec:discussion}). Instead of predicting a single class, the V--A head
produces a probability distribution over a discretized V--A grid: a multi-layer
perceptron maps the fused representation to $81$ logits, which a softmax turns
into a distribution that is reshaped into a $9\times9$ matrix $\mathbf{P}$, with
\begin{equation}
  \mathbf{P}_{ij} = P(\valence = v_i,\ \arousal = a_j \mid x),
  \qquad \sum_{i,j}\mathbf{P}_{ij}=1,
  \label{eq:va-matrix}
\end{equation}
where $\{v_i\}_{i=1}^{9}$ and $\{a_j\}_{j=1}^{9}$ are nine evenly spaced grid
coordinates, $\{1.0, 1.5, \dots, 5.0\}$, placed directly on the native IEMOCAP
valence and arousal annotation scale $[1,5]$ (so no rescaling is needed to
compare predictions with annotations).

\paragraph*{Continuous read-out and uncertainty}
Although the head is trained as a classifier over the $81$ cells, the matrix
$\mathbf{P}$ admits two continuous read-outs used in the analysis of
Sec.~\ref{sec:results-analysis}. Its center of mass is a point estimate of the
affect, on the same $[1,5]$ scale as the ground truth,
\begin{equation}
  \widehat{\valence} = \sum_{i,j} v_i\,\mathbf{P}_{ij}, \qquad
  \widehat{\arousal} = \sum_{i,j} a_j\,\mathbf{P}_{ij},
  \label{eq:va-expectation}
\end{equation}
so the head emits a continuous V--A estimate without any explicit
point-regression loss --- the ground-truth coordinates supervise the head only
through the soft target's location (Eq.~\eqref{eq:soft-target}), not through a
regression objective on $(\widehat{\valence},\widehat{\arousal})$ themselves. The \emph{sharpness} of the
prediction is summarized by the Shannon entropy of the distribution,
\begin{equation}
  H(\mathbf{P}) = -\sum_{i,j} \mathbf{P}_{ij}\ln \mathbf{P}_{ij},
  \quad 0 \le H(\mathbf{P}) \le \ln 81 \approx 4.39\ \text{nats},
  \label{eq:entropy}
\end{equation}
which is maximized by the uniform distribution and which the soft-target width
$\sigma$ (Eq.~\eqref{eq:soft-target}) directly controls.

\paragraph*{Soft target}
The ground-truth annotation $(\valence^\star,\arousal^\star)$ is converted into a
two-dimensional isotropic Gaussian soft target on the same grid,
\begin{equation}
  \mathbf{Q}_{ij} \propto
  \exp\!\left(-\frac{(v_i-\valence^\star)^2 + (a_j-\arousal^\star)^2}{2\sigma^2}\right),
  \qquad \sum_{i,j}\mathbf{Q}_{ij}=1,
  \label{eq:soft-target}
\end{equation}
with $\sigma=0.5$. We emphasize what this target is and is not: it is a
\emph{researcher-chosen spatial smoothing prior} that encodes the assumed
smoothness of the affective space around the (mean) annotation, rather than a
measured, per-utterance annotation distribution --- IEMOCAP's released
dimensional labels are rater averages, and we do not model rater disagreement
here. The chosen width is at least of a plausible scale: parsing the raw
per-evaluator attribute ratings gives a mean within-utterance rater standard
deviation of $0.31$ (valence) and $0.44$ (arousal) on the same $[1,5]$ scale,
so $\sigma=0.5$ is comparable to typical rater spread, though uniform across
utterances. The target therefore prevents the head from being driven toward a
degenerate one-hot spike, but the predicted matrix should be read as the
model's distributional output, not as calibrated annotator uncertainty
(we assess calibration separately in Sec.~\ref{sec:results-analysis}).

\paragraph*{Objective}
The head is trained with a convex combination of cross-entropy on the discrete
emotion label and KL divergence to the soft target,
\begin{equation}
  \mathcal{L} = (1-\lambda)\,\mathcal{L}_{\mathrm{CE}}
              + \lambda\,\mathrm{KL}\!\left(\mathbf{Q}\,\|\,\mathbf{P}\right),
  \qquad \lambda = 0.3 .
  \label{eq:va-loss}
\end{equation}
Here $\mathcal{L}_{\mathrm{CE}}$ is the inverse-frequency class-weighted categorical
cross-entropy on the four-way head (Sec.~\ref{sec:setup-impl}), and
$\mathrm{KL}(\mathbf{Q}\,\|\,\mathbf{P})=\sum_{ij}\mathbf{Q}_{ij}\ln(\mathbf{Q}_{ij}/\mathbf{P}_{ij})$
is taken with the soft target $\mathbf{Q}$ as the reference distribution, so the
gradient pulls the prediction $\mathbf{P}$ toward $\mathbf{Q}$.
\paragraph*{What the two terms do}
The two terms in Eq.~\eqref{eq:va-loss} play complementary roles. The
cross-entropy term acts on the four-way categorical head and anchors the model
to the discrete label, keeping the system comparable to standard SER. The KL term
acts on the $9\times9$ grid and is minimized when $\mathbf{P}=\mathbf{Q}$; because
$\mathbf{Q}$ is a \emph{spread} Gaussian rather than a one-hot vector, the optimum
is itself a distribution, so the head is never driven toward a degenerate spike.
Two design parameters control the shape of that optimum. The width $\sigma$ in
Eq.~\eqref{eq:soft-target} sets how much probability mass the target places off
the ground-truth cell: as $\sigma\!\to\!0$ the target approaches a one-hot vector
and the KL term degenerates to hard cross-entropy on the grid; as $\sigma$ grows
the target flattens and the predicted distribution is pushed to be smoother
(higher entropy). The mixing weight $\lambda$ trades the categorical anchor
against the distributional target. We use $\lambda=0.3$ throughout and treat
$\sigma$ as the principal lever, sweeping it in Sec.~\ref{sec:results-vahead}:
empirically, lowering $\sigma$ sharpens the predictive distribution (its entropy
falls) at no measurable cost in classification accuracy, which is the behavior
Eq.~\eqref{eq:va-loss} predicts.

\paragraph*{Geometry-aware alternative (and why we do not use it)}
The KL term treats the $81$ cells as an unordered set; it does not ``know'' that
adjacent cells are nearby in the V--A plane. A natural alternative is an
optimal-transport cost that penalizes predicted mass by its squared distance to
$(\valence^\star,\arousal^\star)$ --- equivalently the squared $2$-Wasserstein
distance to a Dirac at the label,
\begin{equation}
  W_2^2\big(\mathbf{P},\,\delta_{(\valence^\star,\arousal^\star)}\big)
   = \sum_{ij}\mathbf{P}_{ij}\big[(v_i-\valence^\star)^2+(a_j-\arousal^\star)^2\big],
  \label{eq:w2}
\end{equation}
normalized by $(\valence_{\max}-\valence_{\min})^2+(\arousal_{\max}-\arousal_{\min})^2$
to $[0,1]$. We implemented this and found it a
\emph{negative result}: used alone it tends to collapse mass onto a single cell
and is seed-unstable, and added to KL it does not sharpen the distribution beyond
what $\sigma$ already achieves (Sec.~\ref{sec:results-vahead}). We therefore
retain the KL\,+\,CE objective and report the geometry-aware loss as a controlled
negative.

\section{Experimental Setup}\label{sec:setup}

\subsection{Datasets}\label{sec:setup-data}
All experiments in this paper use \textbf{IEMOCAP}~\cite{Busso2008IEMOCAP},
which provides per-utterance speech, transcripts, categorical labels, and
V--A(--D) values; following common practice we use four classes (\textit{angry};
\textit{happy}+\textit{excited} merged; \textit{sad}; \textit{neutral}), giving
$\sim$$5.5$k utterances over five sessions of distinct speaker pairs. The
released dimensional annotations are rater averages on a $[1,5]$ scale.

The choice of corpus is constrained by the output space: the $9\times9$ V--A
head requires \emph{dimensional} Valence--Arousal annotations, which rules out
sentiment-only corpora (e.g.\ CMU-MOSEI~\cite{Zadeh2018CMUMOSEI},
CH-SIMS~\cite{Yu2020CHSIMS} --- sentiment supplies valence
but not arousal). The natural extensions with native V--A(--D) labels are
MSP-IMPROV~\cite{Busso2017MSPIMPROV} (acted, session structure compatible with
LOSO) and MSP-Podcast~\cite{Lotfian2019MSPPodcast} (large, naturalistic, with a
standard speaker-independent partition). MSP-IMPROV is already exercised here
as a \emph{zero-shot} transfer probe (Sec.~\ref{sec:discussion}, no MSP data
in training); trained extensions to both corpora are future work, requiring
only per-corpus affine normalization of the annotation axes and dropping the
Dominance dimension to share the V--A grid.

\subsection{Evaluation Protocol}\label{sec:setup-protocol}
We adopt a 5-fold leave-one-session-out (LOSO) protocol with rotating-session
inner validation. For fold $f\!\in\!\{0,\dots,4\}$ we set
$\text{test}=\mathcal{S}_f$, $\text{val}=\mathcal{S}_{(f-1)\bmod 5}$, and train on
the remaining three sessions, where $\mathcal{S}_f$ is the $f$-th IEMOCAP session.
Because IEMOCAP sessions correspond to distinct speaker pairs, this yields
\emph{speaker-independent} evaluation. Crucially, early stopping (patience $7$ on
validation UA) and best-checkpoint selection use the \emph{validation} session
only; every headline metric and the reported confusion matrix are computed once
on the untouched test session, so the test set never participates in
\emph{per-run} model selection. We state the study-level caveat explicitly:
because the same five test sessions are evaluated across the backbone, head,
and $\sigma$ ablations reported in this paper, the aggregate comparison reuses
the test partition, as is the norm for benchmark studies; we mitigate this by
fixing the protocol and hyperparameters before the comparisons and by treating
small gaps as noise (Sec.~\ref{sec:results-main}). Our primary metric is
unweighted accuracy (UA, the mean of per-class accuracies, robust to the class
imbalance of IEMOCAP); we also report weighted accuracy (WA) and Macro-F1,
each as the mean$\pm$std over the five test sessions.

\paragraph*{Metrics}
For $C$ classes with per-class recall $\mathrm{acc}_c=\mathrm{TP}_c/N_c$ and
$N=\sum_c N_c$ utterances, unweighted accuracy is the mean per-class recall
$\mathrm{UA}=\frac{1}{C}\sum_{c}\mathrm{acc}_c$, weighted accuracy is the overall
accuracy $\mathrm{WA}=\sum_c (N_c/N)\,\mathrm{acc}_c$, and Macro-F1 is the unweighted
mean of the per-class F1 scores, $\frac{1}{C}\sum_{c}\frac{2\,P_c\,\mathrm{acc}_c}{P_c+\mathrm{acc}_c}$,
with per-class precision $P_c$ (recall being $\mathrm{acc}_c$). For the continuous read-out of the V--A head
(Sec.~\ref{sec:results-analysis}), between the center-of-mass estimate $\hat{z}$
(Eq.~\eqref{eq:va-expectation}) and the ground-truth attribute $z$ (valence or
arousal) we report Pearson $r$, RMSE, MAE on the $[1,5]$ scale, and Lin's
concordance correlation coefficient
\begin{equation}
  \mathrm{CCC} = \frac{2\,\mathrm{cov}(\hat{z},z)}
    {\sigma_{\hat{z}}^2 + \sigma_{z}^2 + (\mu_{\hat{z}}-\mu_{z})^2}.
  \label{eq:ccc}
\end{equation}
Calibration uses the expected calibration error over $B{=}10$ equal-width confidence
bins $\{\mathcal{B}_b\}$,
$\mathrm{ECE}=\sum_{b}\frac{|\mathcal{B}_b|}{N}\big|\mathrm{acc}(\mathcal{B}_b)-\mathrm{conf}(\mathcal{B}_b)\big|$,
and per-utterance distributional sharpness uses the Shannon entropy of
Eq.~\eqref{eq:entropy}.

\subsection{Implementation Details}\label{sec:setup-impl}
The text encoder (BERT-base) is always frozen. The Wav2Vec2 front-end is frozen;
the raw-convolution front-end is pre-trained on LibriSpeech and then either frozen
or fine-tuned (we report both). All other parameters --- projections, temporal
backbone, cross-attention fusion, and head(s) --- are trained. Shared
hyperparameters: shared width $d{=}256$, $6$ backbone layers per modality and
$2$ cross-attention blocks ($8$ heads), dropout $0.3$. Optimization uses AdamW
(base learning rate $1\!\times\!10^{-4}$, weight decay $1\!\times\!10^{-4}$;
parameters the \texttt{mamba\_ssm} library flags as no-decay --- $A$, $D$, and
$\Delta$ biases --- are exempted from weight decay), gradient clipping at
$1.0$, a cosine schedule with linear warmup over the first
$10\%$ of epochs, and bf16 mixed precision (autocast; no loss scaling needed).
Per-category learning-rate multipliers are applied on the base rate: Mamba
parameters $\times3$, encoders $\times0.1$ (only active when fine-tuned),
cross-attention and all others $\times1$. We train for $30$ epochs with early
stopping (patience $7$ on validation UA) and class-weighted cross-entropy,
with inverse-frequency weights computed on each fold's \emph{training split
only} (normalized so the rarest class has weight $1$). The V--A head uses a
$9\times9$ grid over $[1,5]^2$, soft-target width $\sigma{=}0.5$, and loss weight
$\lambda{=}0.3$ (Eq.~\eqref{eq:va-loss}); three IEMOCAP utterances carry a
valence annotation of $5.5$, outside the nominal $[1,5]$ scale, and are clipped
to the grid edge. Audio is capped at $11$\,s (truncating
$\sim$$5\%$ of utterances) and text at $64$ tokens. In the Wav2Vec2 setting
every batch is padded to the full $11$\,s (Sec.~\ref{sec:method-frontend});
the mask-aware raw-conv encoder is padding-invariant, so the long-context
setting uses dynamic padding, with the batch size reduced per model to
fit memory (from $48$ for \texttt{late\_fusion} down to $4$ for the
fine-tuned Transformer); no training batch is dropped (\texttt{drop\_last} is
disabled), so every model sees the identical training set each epoch. Each fold
is seeded deterministically (base seed $42$ plus the fold index, re-applied at
fold start, so single-fold and full runs are bit-identical); the V--A $\sigma$
ablation additionally averages over seeds $\{42,1,2\}$. The headline backbone
table (Table~\ref{tab:main}) likewise averages
over seeds $\{42,1,2\}$: seed $42$ from the primary campaign and
seeds $1,2$ from a follow-up campaign, both under the same pinned software
stack. A separate three-seed rerun of the featured configuration (the
$\sigma$-ablation's fine-tuned $\sigma{=}0.5$ cell,
Table~\ref{tab:va-loss-ablation}) gives $0.721$ UA against
Table~\ref{tab:main}(b)'s $0.730$; run-to-run campaign variance is
therefore $\approx$$0.01$ UA, which is why every comparative claim in this
paper rests on paired, within-fold statistics rather than absolute
third-decimal differences. All training, evaluation, and analysis code was
developed with generative-AI assistance under author verification (see the
declaration before the references) and is released in full (Data and Code
Availability).

Concretely, ``significant'' throughout refers to a two-sided paired $t$-test
over the five fold-level (session-level) deltas, seeds averaged within fold,
reported alongside the \emph{enumerated} session-cluster bootstrap: all
$5^5{=}3125$ with-replacement resamples of the five held-out sessions of the
seed-averaged fold statistic, with $p_{\text{boot}}$ the one-sided share of
resamples at or below zero (``exact'' means exhaustive enumeration, nothing
stronger). Equivalence claims use TOST at
$\alpha{=}0.05$, i.e.\ the $90\%$ CI contained in $\pm0.02$ UA. With
$J{=}5$ clusters, permutation-based cluster-robust alternatives floor at
$p{=}2/2^5{=}0.0625$ two-sided: under the enumerated wild-cluster
(Rademacher) bootstrap-$t$~\cite{Cameron2008Wild} the $+0.016$ dual-head
lift sits exactly at that floor (all five fold deltas positive) and the
dual-head-vs-Transformer contrast at $p{=}0.125$ --- at this fold count
cluster-robust inference is directional support rather than an independent
significance instrument, and we phrase claims accordingly. We run three
primary paired contrasts per operating point, report all of them rather
than selecting, and apply no further multiplicity correction; no conclusion
rests on a single comparison. All runs use a single
NVIDIA GeForce RTX~5090 ($32$\,GB). For reproducibility we pin the software
stack: Python~3.12, PyTorch~2.9.1 (CUDA~12.8), \texttt{transformers}~5.8.0,
\texttt{mamba\_ssm}~2.3.2.post1, and \texttt{causal\_conv1d}~1.6.2.post1.

\section{Results}\label{sec:results}

\subsection{Main Comparison}\label{sec:results-main}
We report the controlled backbone comparison at the two acoustic operating points
of Sec.~\ref{sec:method-frontend}: the SSL front-end (Wav2Vec2, $T\!\approx\!550$,
Table~\ref{tab:main}(a)) and the long-context trainable front-end (raw-conv,
$T\!\approx\!2750$, Table~\ref{tab:main}(b)). Every cell is the mean$\pm$std
over the five held-out LOSO test sessions; within each panel the encoder, fusion,
depth, and width are identical and only the temporal backbone (and the presence
of the V--A head) varies. All numbers come from the rotating-validation protocol
of Sec.~\ref{sec:setup-protocol}.\footnote{The headline backbone tables pool
the primary campaign (seed $42$) with a follow-up multi-seed
campaign (seeds $1,2$) under the same pinned source and configuration files;
the $\sigma$-ablation rows and the
multi-seed entropy--ambiguity checks use designated companion runs under the same
pinned setup (see Sec.~\ref{sec:setup-protocol} for the cross-campaign
replication disclosure).} Results from earlier, weaker protocols are not
comparable and are not reported. \emph{We emphasize that the two panels are not a
clean ``raw-conv vs.\ Wav2Vec2'' comparison}: they differ in both the encoder and
the sequence length, and are presented as two self-contained backbone studies.

\begin{table*}[t]
\centering
\caption{Backbone comparison on IEMOCAP, 5-fold speaker-independent LOSO, at
the two acoustic operating points: \emph{(a)} the frozen \textbf{Wav2Vec2}
front-end ($T\!\approx\!550$); \emph{(b)} the long-context
\textbf{raw-convolution} front-end ($T\!\approx\!2750$, encoder fine-tuned).
UA cells are mean$\pm$std over \textbf{three seeds} $\{42,1,2\}$ with the
per-fold LOSO std in parentheses; WA and Macro-F1 carry the seed std. The
last column is the seed-averaged paired UA difference against the
Transformer fusion row of the same panel with its enumerated session-cluster
bootstrap $95\%$ CI (Sec.~\ref{sec:setup-protocol}). Best UA/WA/Macro-F1 in
bold within each panel. Params = \emph{trainable} parameters in millions: in
\emph{(a)} the Wav2Vec2 ($\sim$95M) and BERT ($\sim$110M) encoders are
frozen and shared across all rows, so only the projection, temporal
backbone, fusion, and head(s) are trained; in \emph{(b)} the fine-tuned
raw-conv encoder ($\sim$1M) is included and the frozen BERT ($\sim$110M) is
excluded. Table body generated from the released statistics files (Data and
Code Availability).}
\label{tab:main}
\small
\begin{tabular}{@{}lccccc@{}}
\toprule
Model & Params & UA $\pm$seed (fold) & WA & Macro-F1 & $\Delta$UA vs.\ \texttt{cross\_attn} [95\% CI] \\
\midrule
\multicolumn{6}{@{}l}{\emph{(a) Frozen Wav2Vec2 front-end ($T\!\approx\!550$)}} \\
\midrule
\texttt{late\_fusion} & $0.46$ & $0.672{\pm}0.003\,(0.017)$ & $0.657{\pm}0.003$ & $0.657{\pm}0.002$ & $-0.007\;[-0.015,+0.001]$ \\
\texttt{cross\_attention} & $13.2$ & $0.679{\pm}0.004\,(0.020)$ & $0.658{\pm}0.004$ & $0.660{\pm}0.005$ & --- \\
\texttt{mamba\_fusion} & $15.8$ & $0.688{\pm}0.003\,(0.030)$ & $0.673{\pm}0.002$ & $0.676{\pm}0.003$ & $+0.009\;[-0.005,+0.023]$ \\
\texttt{mamba\_v2\_fusion} & $16.4$ & $0.684{\pm}0.005\,(0.024)$ & $0.668{\pm}0.008$ & $0.671{\pm}0.008$ & $+0.005\;[-0.003,+0.012]$ \\
\addlinespace
\texttt{mamba\_dual\_head} & $15.9$ & $\mathbf{0.691{\pm}0.004\,(0.018)}$ & $\mathbf{0.678{\pm}0.004}$ & $\mathbf{0.680{\pm}0.004}$ & $+0.012\;[+0.003,+0.022]$ \\
\texttt{mamba\_v2\_dual\_head} & $16.6$ & $0.689{\pm}0.005\,(0.024)$ & $0.671{\pm}0.010$ & $0.674{\pm}0.012$ & $+0.010\;[-0.000,+0.019]$ \\
\midrule
\multicolumn{6}{@{}l}{\emph{(b) Raw-convolution front-end, fine-tuned ($T\!\approx\!2750$)}} \\
\midrule
\texttt{late\_fusion} & $1.45$ & $0.692{\pm}0.007\,(0.037)$ & $0.673{\pm}0.008$ & $0.677{\pm}0.008$ & $-0.008\;[-0.019,+0.000]$ \\
\texttt{cross\_attention} & $14.2$ & $0.700{\pm}0.012\,(0.039)$ & $0.683{\pm}0.013$ & $0.688{\pm}0.014$ & --- \\
\texttt{mamba\_v3\_fusion} & $17.7$ & $0.715{\pm}0.005\,(0.036)$ & $0.700{\pm}0.004$ & $0.703{\pm}0.005$ & $+0.015\;[+0.000,+0.030]$ \\
\texttt{mamba\_v2\_fusion} & $17.4$ & $0.719{\pm}0.008\,(0.036)$ & $0.700{\pm}0.006$ & $0.705{\pm}0.007$ & $+0.020\;[+0.006,+0.033]$ \\
\texttt{mamba\_fusion} & $16.8$ & $0.714{\pm}0.002\,(0.030)$ & $0.697{\pm}0.002$ & $0.700{\pm}0.003$ & $+0.014\;[-0.000,+0.030]$ \\
\addlinespace
\texttt{mamba\_v2\_dual\_head} & $17.6$ & $0.723{\pm}0.006\,(0.039)$ & $0.707{\pm}0.007$ & $0.710{\pm}0.007$ & $+0.023\;[+0.007,+0.035]$ \\
\texttt{mamba\_dual\_head} & $16.9$ & $\mathbf{0.730{\pm}0.003\,(0.037)}$ & $\mathbf{0.714{\pm}0.003}$ & $\mathbf{0.718{\pm}0.002}$ & $+0.030\;[+0.010,+0.047]$ \\
\bottomrule
\end{tabular}

\end{table*}

Three findings are consistent across both front-ends. \emph{(i)} The state-space
backbones (Mamba v1/v2) sit above the Transformer fusion at matched depth
and width: at the short operating point \texttt{mamba\_fusion} ($0.688$ UA)
exceeds \texttt{cross\_attention} ($0.679$), and at the long operating
point the single-head Mamba backbones lead the Transformer by $1.4$--$2.0$ UA
points ($0.714$--$0.719$ vs.\ $0.700$), with the dual-head variants extending
the margin to $2.3$--$3.0$ points. \emph{(ii)} For the \emph{single-head}
backbones this margin is not statistically established: at the short operating
point it is within noise under both of our paired tests, and at the long
operating point it is supported by the enumerated session-cluster bootstrap
($p\!\le\!0.03$) but not by the paired $t$-test ($p\!\ge\!0.07$), with the
per-fold standard deviation ($\sim$$0.03$--$0.04$) comparable to the margin.
The featured \emph{dual-head} systems, however, exceed the Transformer
baseline significantly under \emph{both} tests at the long operating point
(\texttt{mamba\_dual\_head}$\,-\,$\texttt{cross\_attention}: $+0.030$ UA,
paired $t$ $p{=}.044$, enumerated session-cluster bootstrap $p{<}.001$,
$d_z{=}1.29$; \texttt{mamba\_v2\_dual\_head}: $+0.023$, $p_t{=}.047$) --- a
system-level comparison that includes the V--A head, not a pure backbone
contrast. The comparison is also not
parameter-matched: at equal depth and width the bidirectional Mamba blocks carry
$\approx$$2.6$--$3.2$M more trainable parameters than the Transformer
encoder (an $\approx$$18$--$24\%$ increase; Table~\ref{tab:main},
Params column), so the state-space edge is not capacity-free. \emph{(iii)} Mamba-3 is
competitive but not the best: \texttt{mamba\_v3\_fusion} ($0.715$ UA) sits
mid-pack --- above the Transformer and on par with Mamba-1, but below Mamba-2 ---
while carrying the highest long-sequence latency and memory
(Sec.~\ref{sec:results-cost}); it thus offers neither an accuracy nor an
efficiency reason to prefer it here, as we discuss in Sec.~\ref{sec:discussion}.

\subsection{Encoder Pre-Training Ablation}\label{sec:results-encpretrain}
The raw-convolution front-end is pre-trained on LibriSpeech
(Sec.~\ref{sec:method-frontend}); we pre-trained it two ways and ask whether the
choice transfers downstream. The \texttt{basic} variant pre-trains the
convolutional stack alone by self-supervised mel reconstruction, whereas the
\texttt{mamba} variant attaches a Mamba-2 context module during pre-training (the
context module is discarded before fine-tuning, so the downstream encoder is the
\emph{identical} conv-only stack in both cases). The context module markedly
lowers the \emph{pre-training} objective, but Table~\ref{tab:enc-pretrain} shows
this does \emph{not} carry over to emotion recognition: across all seven backbones
the two variants are indistinguishable within the per-fold standard deviation
($\sim$$0.03$--$0.04$ UA). We therefore report the \texttt{basic} arm throughout;
it is within fold std of the \texttt{mamba} arm on every model and, at the
ablation's fixed seed, contains the top raw-conv configurations
(\texttt{mamba\_v2\_fusion} $0.728$ and \texttt{mamba\_dual\_head} $0.727$ UA,
indistinguishable within fold std; the three-seed means of
Table~\ref{tab:main}(b) confirm the dual-head arm at $0.730$). The
result also suggests that the learned convolutional filters, rather than the
pre-training context module, are what the downstream model relies on.

\begin{table}[t]
\centering
\caption{Encoder pre-training ablation: UA (mean$\pm$std over the five LOSO test
sessions, single seed 42) for the two LibriSpeech pre-training variants of the
raw-convolution front-end ($T\!\approx\!2750$, fine-tuned). \texttt{basic}: conv-only SSL;
\texttt{mamba}: SSL assisted by a Mamba-2 context module that is stripped before
fine-tuning. The two are within fold std on every model.}
\label{tab:enc-pretrain}
\begin{tabular}{lcc}
\toprule
Model & \texttt{basic} UA & \texttt{mamba} UA \\
\midrule
\texttt{late\_fusion}          & $0.695{\pm}0.039$ & $0.678{\pm}0.029$ \\
\texttt{cross\_attention}      & $0.688{\pm}0.039$ & $0.708{\pm}0.028$ \\
\texttt{mamba\_v3\_fusion}     & $0.720{\pm}0.030$ & $0.718{\pm}0.031$ \\
\texttt{mamba\_v2\_fusion}     & $\mathbf{0.728{\pm}0.034}$ & $0.717{\pm}0.033$ \\
\texttt{mamba\_fusion}         & $0.714{\pm}0.033$ & $0.714{\pm}0.019$ \\
\texttt{mamba\_v2\_dual\_head} & $0.725{\pm}0.035$ & $0.727{\pm}0.024$ \\
\texttt{mamba\_dual\_head}     & $0.727{\pm}0.040$ & $0.724{\pm}0.038$ \\
\bottomrule
\end{tabular}
\end{table}

\subsection{Effect of the Probabilistic V--A Head}\label{sec:results-vahead}
Adding the $9\times9$ V--A head is the paper's central design choice, and a
prerequisite for it is that the extra objective must not harm the categorical
task it is bolted onto. It does not --- it helps. Comparing each
\texttt{*\_dual\_head} model against its single-head counterpart under matched
training, the dual head never hurts and usually helps: with Wav2Vec2 it lifts
\texttt{mamba\_fusion} from $0.688$ to $0.691$ UA (three-seed means) --- the top
configuration at this operating point --- and with the raw-conv front-end it
lifts \texttt{mamba\_fusion} from $0.714$ to $0.730$ UA, \emph{above} the
strongest single-head backbone (\texttt{mamba\_v2\_fusion}, $0.719$); the
seed-averaged paired lift is $+0.016$ UA with a fold-level $95\%$ CI of
$[+0.001, +0.030]$ (all five fold deltas positive; note the lift is smaller
than the $\pm0.02$ TOST margin used for equivalence claims, so a contrast
can be simultaneously significant and equivalence-bounded --- we report
both), and the corresponding Mamba-2 pair is equivalent within
$\pm 0.02$ by TOST. Across the dual-vs-single pairs spanning
encoders and freeze settings the dual head improves UA or matches it within the
per-fold standard deviation. It does so at a negligible parameter cost: the
$9\times9$ head adds only about $0.15$M trainable parameters over the single-head
model (Table~\ref{tab:main}), so the distributional
output comes essentially for free. The categorical anchor (the cross-entropy term) and
the distributional target (the KL term) are thus complementary: the soft-label
head acts as an auxiliary objective that preserves --- and at both operating
points \emph{improves} --- classification while, as a by-product, yielding the
full V--A distribution that the rest of this section analyzes. We scope the
attribution honestly: the pre-specified null controls of
Sec.~\ref{sec:results-controls} show that simpler V--A auxiliaries (a scalar
regression head, two factorized marginal heads) reproduce this classification
lift within noise, so the lift is credited to V--A auxiliary supervision in
general --- the \emph{distributional} form is justified by the distribution it
yields, not by extra accuracy.

\subsection{Analysis of the V--A Probability Matrix}\label{sec:results-analysis}
Beyond classification accuracy, the value of the probabilistic head lies in the
quality of the distribution it produces over the V--A plane. We analyze three
properties on the held-out LOSO test sessions: (i) how well the expected value
(center of mass) of the predicted matrix tracks the continuous ground-truth
attributes; (ii) the sharpness of the per-utterance distribution, measured by
its entropy; and (iii) the calibration of the classifier confidence.

\begin{figure}[t]
  \centering
  \includegraphics[width=\linewidth]{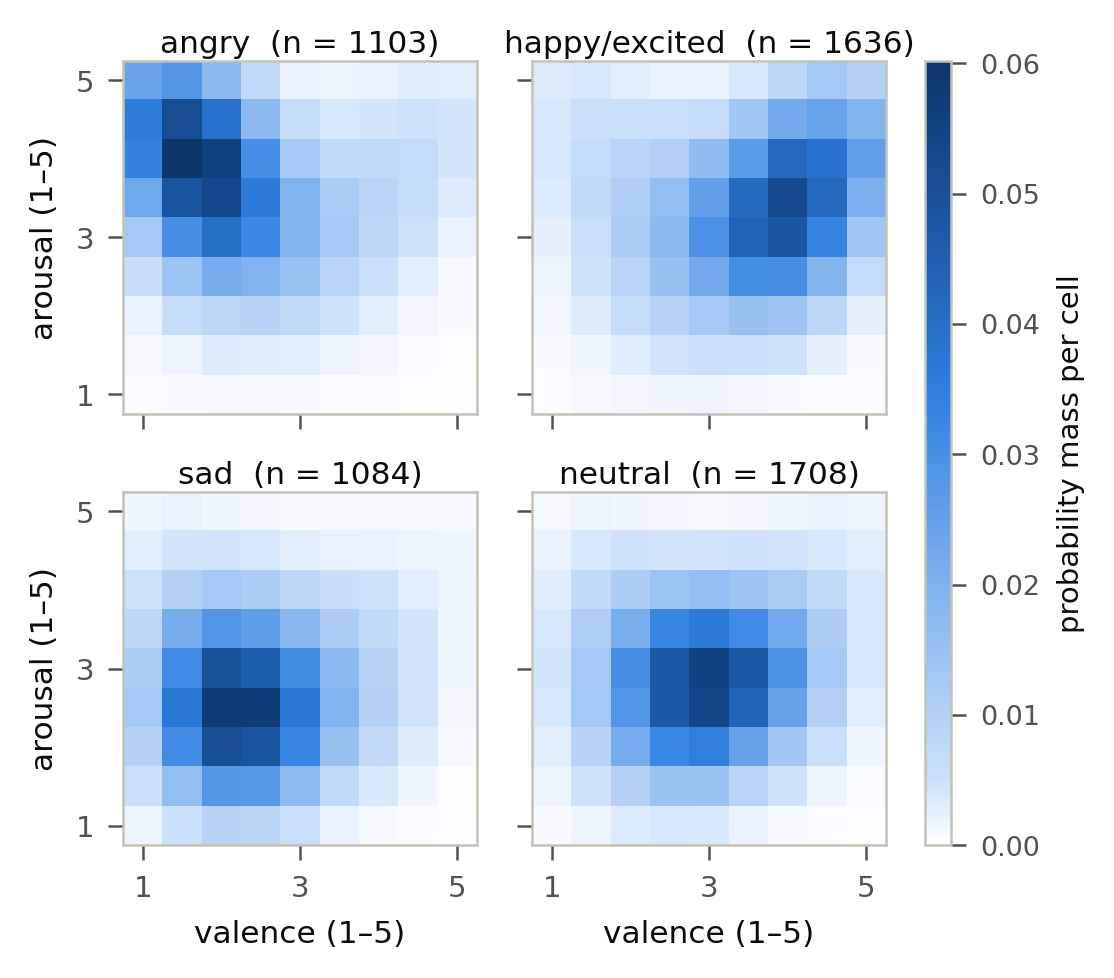}
  \caption{Per-class mean of the predicted $9\times9$ V--A matrix on the
  held-out LOSO test sessions (\texttt{mamba\_dual\_head}, raw-conv encoder;
  representative seed of the three-seed replication; shared color scale).
  Each emotion's mass concentrates in its expected circumplex
  quadrant~\cite{Russell1980Circumplex}: the head has learned class-level
  affective \emph{geometry}; whether it also tracks V--A variation
  \emph{within} a class is answered in Fig.~\ref{fig:va-within}.}
  \label{fig:va-heatmaps}
\end{figure}

\begin{figure}[t]
  \centering
  \includegraphics[width=\linewidth]{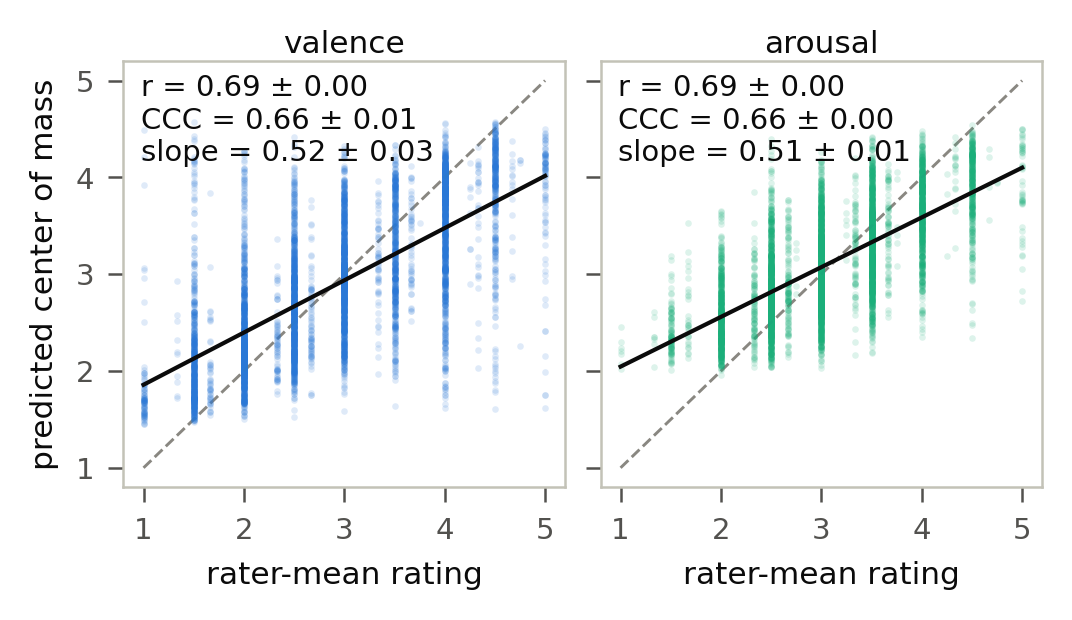}
  \caption{Expected value (center of mass) of the predicted V--A matrix vs.\
  ground-truth valence and arousal (points: representative seed; annotations:
  three-seed mean$\pm$std of $r$, CCC, and the regression slope), obtained
  without a point-regression loss (the coordinates supervise only the soft
  target's location). Vertical striping reflects the discrete support of
  rater-averaged annotations; the fitted slope of $\approx\!0.5$ (solid line
  vs.\ the dashed identity) quantifies the expected shrinkage of a
  center-of-mass read-out on a bounded grid.}
  \label{fig:va-scatter}
\end{figure}

\begin{figure}[t]
  \centering
  \includegraphics[width=\linewidth]{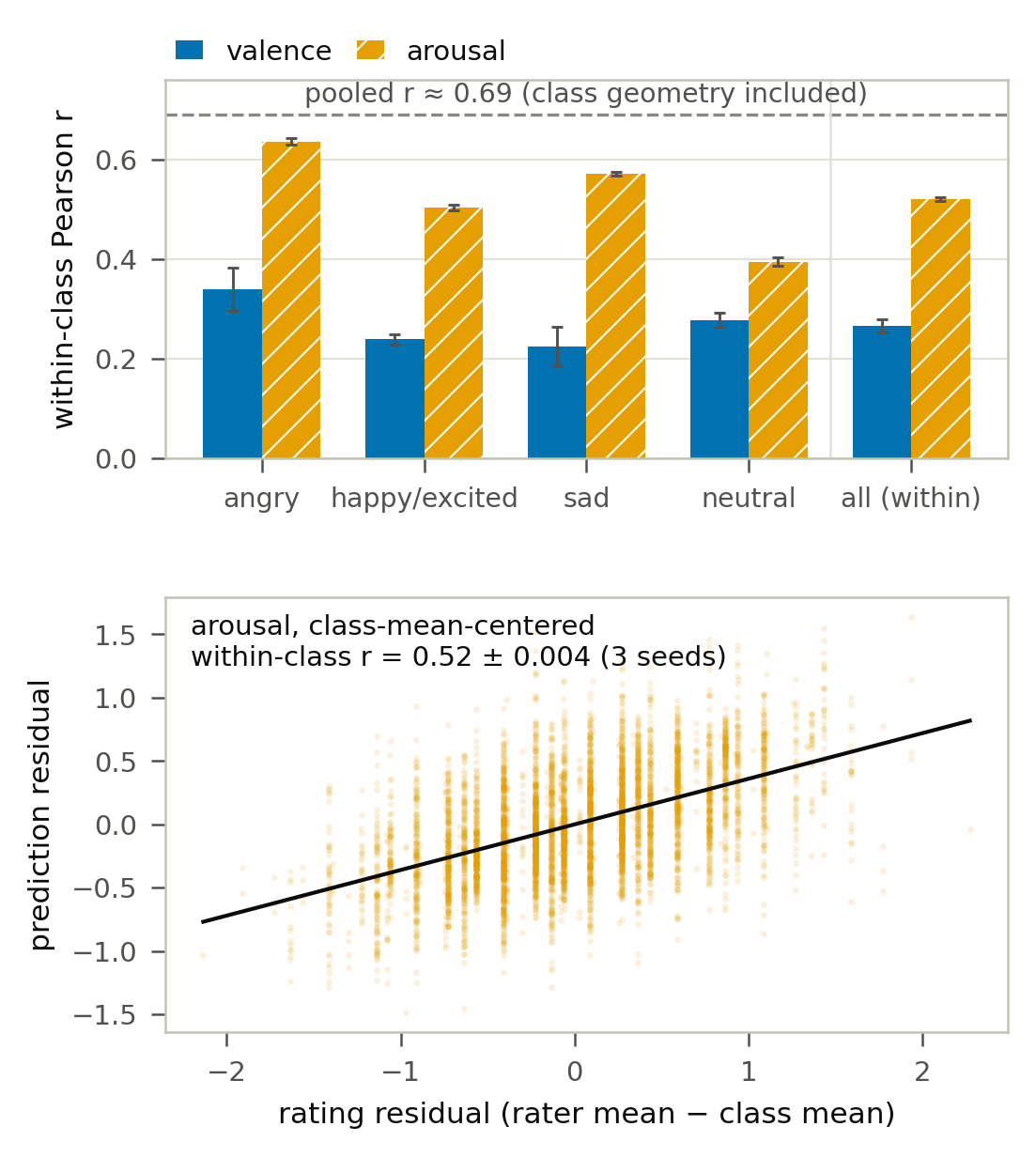}
  \caption{Within-class tracking of the continuous ratings
  (\texttt{mamba\_dual\_head}, raw-conv encoder). Top: class-mean-centered
  Pearson $r$ per emotion and over all classes (bars: three-seed mean; error
  bars: seed std). The pooled correlation (dashed) includes between-class
  geometry; the within-class bars remove it. Bottom: class-mean-centered
  arousal residuals for the representative seed --- within-class arousal
  tracking is genuine ($r=0.52$), while valence retains only $r=0.27$.}
  \label{fig:va-within}
\end{figure}

\paragraph*{The distribution recovers affective geometry}
The per-class mean of the predicted matrices lands in the expected quadrant of
the circumplex~\cite{Russell1980Circumplex} (anger: low $V$/high $A$; happiness:
high $V$/high $A$; sadness:
low $V$/low $A$; neutral: center; Fig.~\ref{fig:va-heatmaps}), and the center of
mass tracks the continuous labels with Pearson $r=0.69$ for valence and
$r=0.69$ for arousal (concordance CCC $0.66$ and $0.66$; RMSE $0.69$ and $0.55$,
MAE $0.52$ and $0.43$ on the $[1,5]$ scale; three-seed means over the
replication runs of Table~\ref{tab:main}(b), pooled over $5{,}531$ test
utterances per seed; Fig.~\ref{fig:va-scatter}).
The head emits this continuous estimate without any explicit
coordinate-regression objective: the KL target supervises the full
distribution (its location built from the rater-mean coordinates,
Eq.~\eqref{eq:soft-target}), and the expectation read-out follows from it.
Two qualifications frame these numbers. First, they are
\emph{pooled} correlations: the four classes occupy distinct circumplex
regions, so between-class separation contributes to the pooled $r$
(replacing every prediction and label by its class mean alone already yields
$r\!\approx\!0.99$). Decomposing the read-out \emph{within} classes answers
the stricter test directly (Fig.~\ref{fig:va-within}): on class-mean-centered
residuals the arousal
read-out still tracks the ratings at $r=0.52$ (three-seed mean; session-level
bootstrap sensitivity $[0.50,0.55]$; per class, anger $0.64$, sadness $0.57$,
happiness $0.50$, neutral $0.40$), whereas valence retains only $r=0.27$
$[0.23,0.32]$. The pooled valence correlation is therefore mostly class
geometry, while within-class arousal tracking is genuine --- consistent with
the broader finding that acoustics carry arousal more readily than valence.
Second, the CCC is now anchored in-pipeline: replacing the $9\times9$ head
with a direct $(v,a)$-regression head under otherwise identical training
(Sec.~\ref{sec:results-controls}, Table~\ref{tab:controls}) reaches CCC
$0.687/0.678$ --- about $0.02$ \emph{above} the distributional read-out's
$0.661/0.661$. The center-of-mass estimate therefore pays a small,
quantified tracking price relative to a dedicated regressor
(cf.\ multitask dimensional SER~\cite{Atmaja2020Multitask}); it is the price
of obtaining the full distribution rather than a point.

\paragraph*{The soft-target width $\sigma$ controls distributional sharpness}
\label{par:sigma-ablation}
We treat the Gaussian soft-target width $\sigma$ of Eq.~\eqref{eq:soft-target}
as the principal lever on the predictive distribution and ablate it under both a
frozen and a fine-tuned acoustic encoder (Table~\ref{tab:va-loss-ablation}).
Sharpening the target from $\sigma{=}0.5$ to $\sigma{=}0.25$ reduces the mean
per-utterance entropy from $3.40$ to $2.97$ nats (from $77\%$ to $68\%$ of the
maximum $\ln 81$; equivalently $3.49\!\to\!3.07$ with a frozen encoder)
\emph{without} any significant change in classification UA
(differences are within the seed/fold standard deviation of $\sim\!0.03$) and
without degrading the center-of-mass regression. This effect is stable across
the frozen and fine-tuned settings, indicating that $\sigma$ offers a
controllable, essentially cost-free knob on the sharpness of the predicted
V--A distribution. Extending the sweep to $\sigma\!\in\!\{0.125, 1.0\}$ leaves UA
within the same per-fold standard deviation across the full range, so $\sigma$
acts as a sharpness control rather than an accuracy lever and the $\sigma{=}0.5$
default is not a tuned choice. We additionally explored a geometry-aware loss that penalizes
predicted mass by its squared V--A distance to the ground-truth point
(equivalently the squared 2-Wasserstein distance $W_2^2$ to a Dirac), used alone
and added to the KL term; neither
variant improved UA nor sharpened the distribution beyond the $\sigma$ sweep, so
we retain the KL\,+\,CE objective and report the geometry-aware loss as a
negative result.

\begin{table}[t]
\centering
\caption{Effect of the soft-target width $\sigma$ on the V--A head
(IEMOCAP, \texttt{mamba\_dual\_head}, raw-conv encoder). UA is
mean over 3 seeds $\times$ 5 LOSO folds; entropy $H$ (in nats, max
$\ln 81\!=\!4.39$) and ECE are pooled over the test sessions of a representative
seed. Lower $\sigma$ sharpens the distribution ($H\!\downarrow$) at no UA cost;
calibration is unaffected.}
\label{tab:va-loss-ablation}
\begin{tabular}{llccc}
\toprule
Encoder & $\sigma$ & UA & $H$ (nats) & ECE \\
\midrule
Frozen      & $0.50$ & $0.726$ & $3.49$ & $0.20$ \\
Frozen      & $0.25$ & $0.730$ & $3.07$ & $0.19$ \\
Fine-tuned  & $0.50$ & $0.721$ & $3.40$ & $0.23$ \\
Fine-tuned  & $0.25$ & $0.727$ & $2.97$ & $0.21$ \\
\bottomrule
\end{tabular}
\end{table}

\begin{figure}[t]
  \centering
  \includegraphics[width=\linewidth]{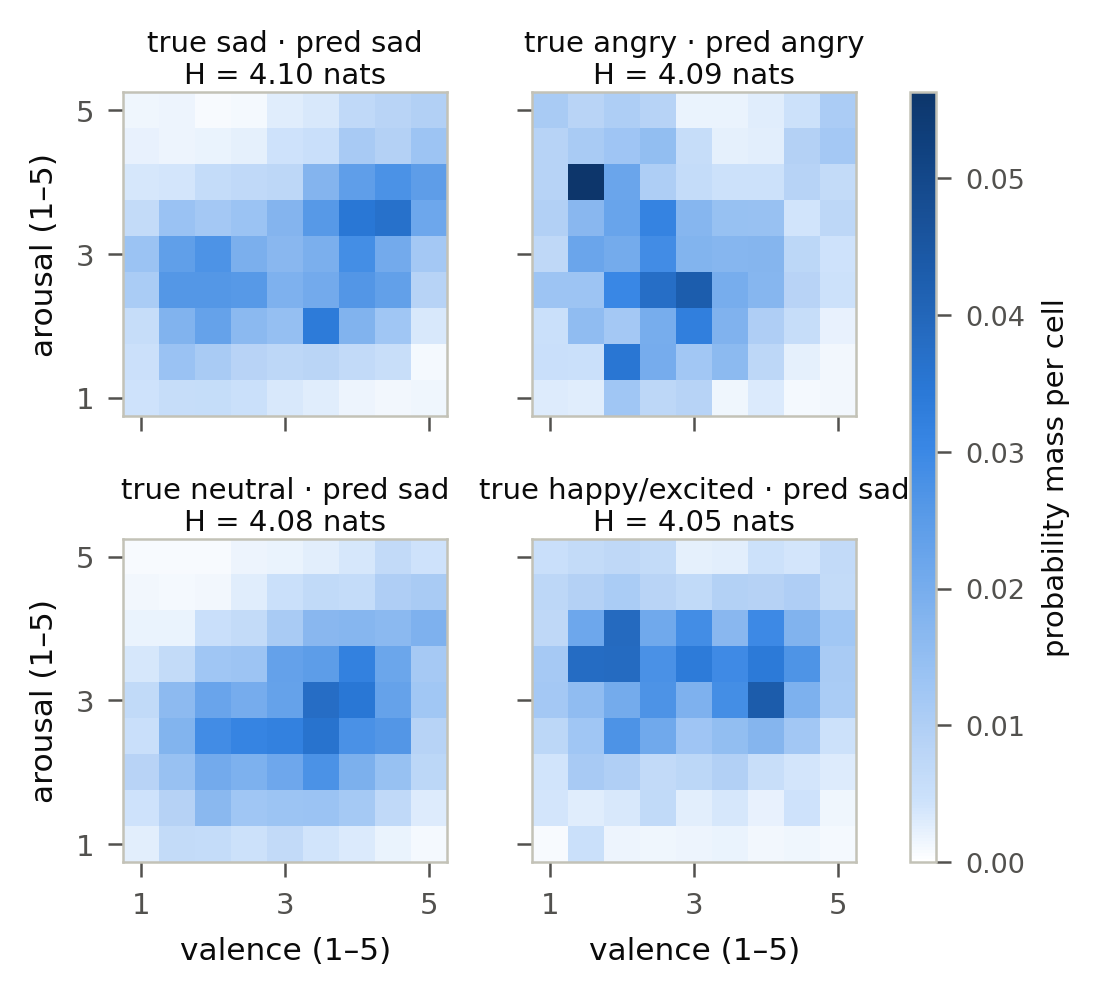}
  \caption{Predicted $9\times9$ V--A matrices for four of the highest-entropy
  ($H\!\approx\!4.1$ nats; max $\ln 81\!=\!4.39$) test utterances
  (\texttt{mamba\_dual\_head}, raw-conv encoder). The mass spreads across
  multiple V--A regions and straddles the true/predicted category boundary ---
  diffuse, multi-peaked model outputs that the grid head can represent and a
  hard classifier cannot.}
  \label{fig:va-examples}
\end{figure}

\paragraph*{What high-entropy predictions look like}
Figure~\ref{fig:va-examples} shows the predicted
$9\times9$ matrices for the highest-entropy test utterances ($H\!\approx\!4.1$
nats, near the maximum $\ln 81\!=\!4.39$). Instead of collapsing onto a single
cell, the mass spreads across adjacent --- and sometimes non-adjacent --- regions
of the V--A plane and straddles the boundary between the true and the confused
category (e.g.\ happy/excited mass leaking toward the high-arousal anger region,
or toward low-valence sadness). We describe these as \emph{diffuse, multi-modal model outputs}, and ask whether
that diffuseness tracks \emph{human} ambiguity. IEMOCAP's per-rater files expose
two distinct notions of ambiguity, and the predicted entropy behaves differently
on each; evaluating a recognizer against inter-labeler consistency in this way
follows a long-standing precedent~\cite{Steidl2005Measure}. Against
\emph{dimensional} disagreement --- the per-rater spread of the
continuous V--A ratings --- entropy shows a \emph{marginally negative}
association across the
$5{,}531$ test utterances (Spearman $-0.06$, session bootstrap $[-0.09,-0.00]$,
an interval that sits at or below zero;
partial correlation given distance-to-grid-edge $-0.03$); on IEMOCAP this
attribute spread rests on only two to three evaluators and sits near a rater
noise floor, which attenuates any observable correlation in either direction. Against \emph{categorical} ambiguity --- the Shannon entropy of the
per-utterance emotion-\emph{category} vote distribution --- predicted entropy is
instead \emph{weakly but consistently positive}: Pearson $\approx\!0.09$, and the
association \emph{survives} the grid-edge control (partial correlation given
distance-to-edge $\approx\!0.07$, within-edge-bin mean $\approx\!0.07$) and is
positive in every one of three seeds ($0.09\!\pm\!0.03$). Entropy also rises with
grid-edge distance ($r\!\approx\!0.25$), the bounded-grid artifact of
Sec.~\ref{sec:discussion}, which is precisely why the categorical link is
reported as an edge-controlled partial correlation. The effect is small
($r^2\!<\!1\%$, in line with IEMOCAP's limited rater pool), and we make no claim
that the entropy is a calibrated ambiguity estimate; but it is a real,
confound-controlled signal that the head's diffuseness carries a modest trace of
which utterances annotators found categorically hard, rather than being a pure
grid artifact. The figure thus shows the output space \emph{can} represent
multi-peaked affect; the correlation analysis shows that this diffuseness is
weakly but genuinely linked to categorical rater ambiguity, while showing a
marginally negative association with dimensional V--A disagreement.

A vote-conditioned placement test sharpens this picture from \emph{how much}
diffuseness to \emph{where} it points --- and surrogate nulls then show the
effect is carried by predicted \emph{location}, not by the joint-pmf format.
For utterances whose consensus label is $c_1$ but where at least one rater
voted a second category $c_2$, the predicted mass inside $c_2$'s circumplex
region (cells within $0.75$ rating units of the data-derived class anchor)
exceeds that of unanimous-$c_1$ controls by $+0.077\pm0.005$ absolute mass
(session-bootstrap sensitivity $[+0.058,+0.092]$; direction-consistent at
radii $0.5$--$1.0$ and across all sufficiently populated label pairs). The
same statistic on the $\sigma{=}0.5$ \emph{training-target} Gaussians built
from the consensus coordinates gives $+0.114$ --- the ceiling inherited from
the raters' own consensus geometry. Three surrogates, scored with the
identical statistic, grouping, and bootstrap, decide what the $+0.077$ is
worth: a $\sigma{=}0.5$ Gaussian placed at the scalar \emph{regression}
head's $(\hat v,\hat a)$ reaches $+0.114$ $[+0.093,+0.135]$ --- the full
ceiling; the classifier posterior mixed over anchor Gaussians
$\sum_c p(c\,|\,x)\,\mathcal{N}(\text{anchor}_c,\sigma)$ reaches $+0.101$
$[+0.075,+0.120]$; both \emph{exceed} the joint head. The pmf does beat the
mixture over its own per-class \emph{mean} matrices ($+0.032$
$[+0.025,+0.038]$), so its placement is utterance-specific rather than a
class template --- but any location-faithful surrogate reproduces the shift,
so the placement channel provides no measured value beyond what a scalar
location model already carries. This is consistent with the regression
anchor's slightly better point tracking
(Sec.~\ref{sec:results-controls}). Alongside, a four-way softmax-entropy
null shows the $9\times9$ entropy adds essentially nothing \emph{beyond} the
classifier's own uncertainty for disagreement \emph{magnitude} (partial
correlation with categorical rater entropy given softmax entropy
$\approx\!0.05$). Together, both ambiguity channels --- spread and placement
--- are explained by simpler surrogates; the head's distributional value on
this corpus lies in the output \emph{format} (one normalized joint pmf with
free continuous read-outs), not in a measured placement advantage.

\paragraph*{How joint is the learned distribution?}
The $81$-cell parameterization \emph{can} express dependence between valence and
arousal, but on IEMOCAP the learned distributions largely do not exercise this
capacity: the per-utterance KL divergence between $\mathbf{P}$ and the outer
product of its own marginals averages $0.016$ nats (90th percentile $0.029$; cf.\
the $\ln 81\!\approx\!4.39$-nat entropy scale), i.e., the predicted joint is
nearly separable for most utterances. This measurement locates the
contribution precisely: on this corpus, with unimodal Gaussian targets, the
joint head's value is its \emph{format} (a single normalized affect
distribution with free continuous read-outs), not measured V--A dependence
--- a distinction we state rather than blur. Whether task pressure that explicitly requires cross-dimension
structure induces non-separable predictions is an open question for the
multi-plane extension we pursue in follow-up work.

\begin{figure}[t]
  \centering
  \includegraphics[width=\linewidth]{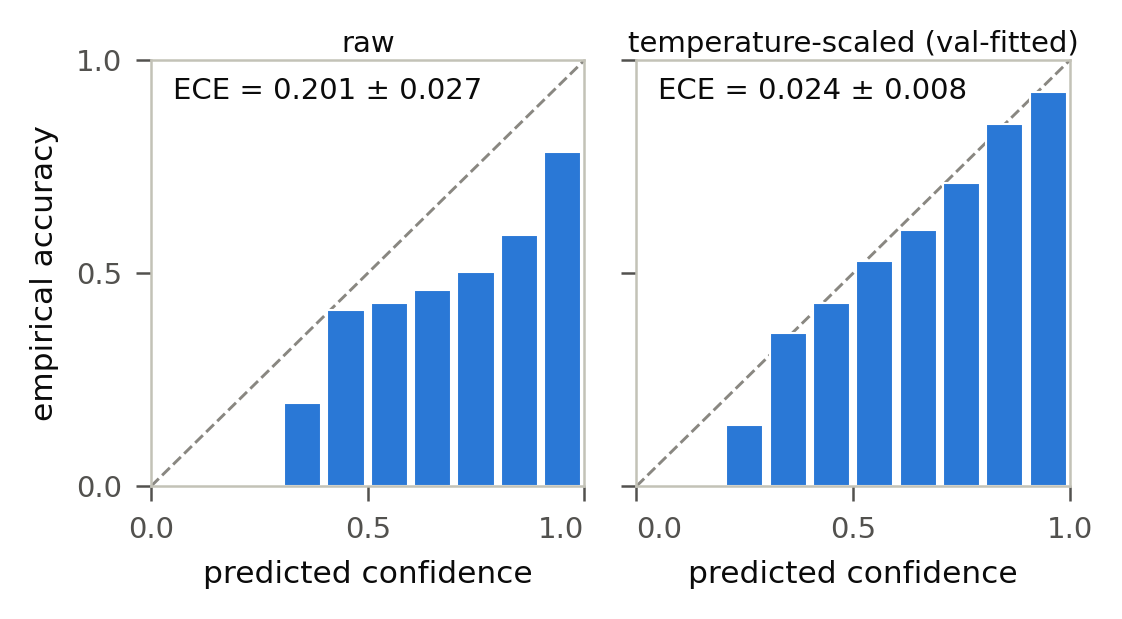}
  \caption{Reliability diagram on the pooled LOSO test sessions (three seeds
  pooled; ECE annotations are three-seed mean$\pm$std). Left: raw confidence
  sits below the diagonal (ECE $0.20$) --- the categorical head is
  over-confident. Right: a scalar temperature fitted per fold on the rotating
  \emph{validation} session only repairs the test ECE to $0.02$ without
  changing any argmax decision.}
  \label{fig:va-calib}
\end{figure}

\begin{figure}[t]
  \centering
  \includegraphics[width=\linewidth]{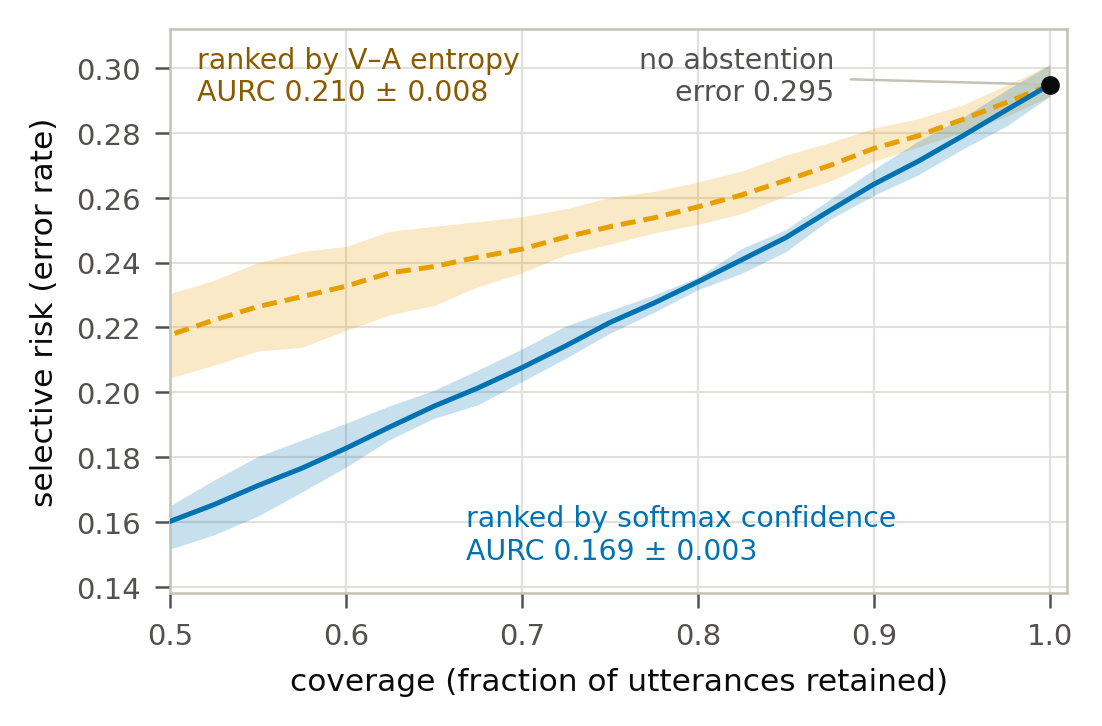}
  \caption{Selective risk vs.\ coverage when abstaining on the least-confident
  utterances, ranked either by softmax confidence or by V--A entropy (lines:
  three-seed mean; bands: seed range). Confidence ranks errors better at every
  coverage --- the plain reading is that the $9\times9$ entropy adds no
  abstention value beyond the classifier's own confidence, even though
  abstention itself helps under either ranking.}
  \label{fig:va-risk}
\end{figure}

\begin{figure}[t]
  \centering
  \includegraphics[width=\linewidth]{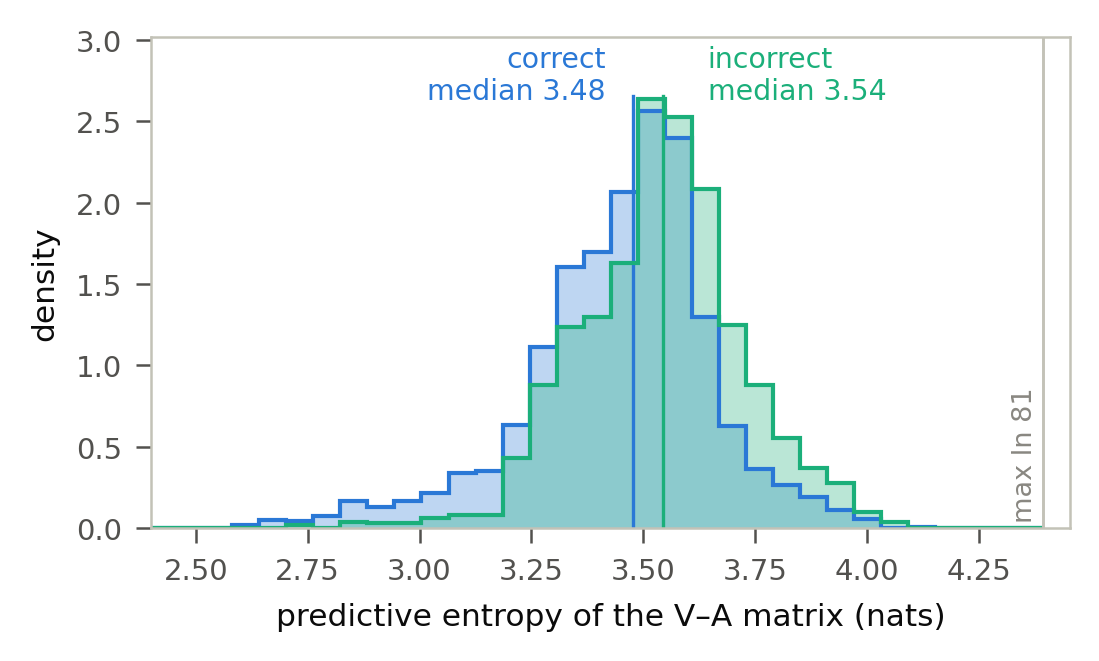}
  \caption{Predictive entropy of the V--A matrix for correct vs.\ incorrect test
  predictions (\texttt{mamba\_dual\_head}, raw-conv encoder; dashed lines mark the
  group centers). The two distributions overlap almost entirely, so entropy is
  only a weak indicator of correctness --- consistent with the over-confidence in
  Fig.~\ref{fig:va-calib}. A high-entropy prediction reflects mass spread over
  the V--A grid (Fig.~\ref{fig:va-examples}), but is not a reliable indicator of
  an incorrect prediction --- nor of rater-measured ambiguity (see text).}
  \label{fig:va-entropy}
\end{figure}

\paragraph*{Calibration}
The classifier is over-confident: the reliability diagram
(Fig.~\ref{fig:va-calib}) lies below the diagonal with an expected calibration
error of ECE~$\approx\!0.20$ (three-seed mean; $0.23$ in the $\sigma$-ablation
campaign, Table~\ref{tab:va-loss-ablation}), essentially unchanged by $\sigma$.
The fix is post-hoc and validation-legal: a scalar temperature fitted per fold
on the rotating \emph{validation} session only (mean $T\!\approx\!2.7$)
repairs the test ECE to $\approx\!0.02$ without changing any argmax decision.
The V--A head errs in the \emph{opposite} direction --- its distributions are
conservatively diffuse --- and the analogous validation-fitted temperature
($T\!\approx\!0.74$, i.e.\ sharpening) brings the highest-density-region
coverage of $\mathbf{P}$ close to nominal across levels (empirical coverage
$0.55$ at the $0.5$ level and $0.90$ at $0.9$, from raw over-coverage of
$0.66$ and $0.95$).
The raw over-confidence also limits the \emph{usefulness} of entropy as a correctness
signal: the predictive entropy of correct and incorrect predictions is nearly
identical (Fig.~\ref{fig:va-entropy}; medians $3.48$ vs.\ $3.54$ nats out of a
maximum $4.39$), so entropy alone barely separates the two. The selective-risk
view (Fig.~\ref{fig:va-risk}) makes the deployment consequence explicit:
ranking by softmax confidence orders errors better than the V--A entropy does
(area under the risk--coverage curve $0.169\pm0.003$ vs.\ $0.210\pm0.008$,
three seeds), though abstaining on the $20\%$ most diffuse predictions still
lifts accuracy \emph{on the retained utterances} from $0.705$ (the pooled
no-abstention accuracy of Fig.~\ref{fig:va-risk}) to $0.743$ --- a selective
accuracy, not a system-level UA. The head therefore
yields a faithful affect \emph{distribution} but is not, as-is, a reliable
confidence or abstention signal --- a limitation we attribute to the same
miscalibration and flag as future work rather than a solved problem.

Throughout this subsection UA and every annotated distribution metric ($r$,
CCC, ECE, AURC, the within-class correlations) are three-seed means over the
replication runs of Table~\ref{tab:main}(b), computed from one committed
per-utterance prediction cache; panels that display individual predictions
(Figs.~\ref{fig:va-heatmaps}, \ref{fig:va-scatter}, \ref{fig:va-examples},
\ref{fig:va-entropy}, and the lower panel of Fig.~\ref{fig:va-within}) show
the representative seed~42 and say so. The entropy--ambiguity correlations
are additionally verified to be sign-stable across the three seeds.

\subsection{Pre-Specified Controls and Extensions}\label{sec:results-controls}
The revision plan pre-specified a set of control and extension arms in dated
internal plan files fixed before any of these runs were launched
(each 3 seeds $\times$ 5 LOSO folds under the featured recipe);
Table~\ref{tab:controls} reports all of them against the featured system,
with seed-averaged paired $\Delta$UA and enumerated session-cluster bootstrap CIs.

\paragraph*{Null-model controls}
Swapping the $9\times9$ head for a scalar $(v,a)$-regression auxiliary
($\Delta$UA $-0.007$ $[-0.019,+0.003]$) or for two factorized $9$-way
marginal heads ($-0.005$ $[-0.016,+0.006]$; equivalent within $\pm0.02$ by
TOST) leaves classification statistically unchanged --- the lift of
Sec.~\ref{sec:results-vahead} comes from V--A auxiliary supervision, not from
the joint grid. This is exactly what the near-separability of the learned
joint (Sec.~\ref{sec:results-analysis}, $0.016$ nats) predicts. On the
continuous read-out the ordering reverses in favor of the simpler heads: the
direct regression anchor reaches CCC $0.687/0.678$ and the marginal heads
$0.673/0.675$, versus $0.661/0.661$ for the joint head's center of mass. The
joint distributional form is therefore \emph{not} the best point-tracker ---
its case rests on what only it provides: a single normalized joint pmf as an
output format behind the coverage and calibration analyses of
Sec.~\ref{sec:results-analysis} (the vote-conditioned placement channel is
itself location-driven --- the surrogate tests there show the same scalar
regression head recovers the full placement ceiling).

\paragraph*{Anisotropic soft-target width}
Setting the per-axis target widths to the measured rater standard deviations
($\sigma_V{=}0.31$, $\sigma_A{=}0.44$) changes neither UA ($-0.002$,
TOST-equivalent) nor the read-out CCC ($0.668/0.665$): within this recipe the
isotropic $\sigma{=}0.5$ default is not a liability, and we report the
rater-matched variant as a clean negative rather than a tuned improvement.

\paragraph*{Rater-vote mixture supervision (pilot)}
Replacing the single Gaussian target with a vote-weighted mixture (anchor
Gaussian at the consensus coordinates plus vote-mass components at the other
voted categories' prototypes) matches UA ($0.728\pm0.007$) but \emph{costs}
arousal tracking (CCC $0.609$ vs.\ $0.661$; valence $0.654$): distributing
target mass across categories dilutes the location signal on the dimension
acoustics carry best. We report this pilot as a negative result for
plain mixture supervision at this scale.

\paragraph*{Modality ablation}
Single-branch baselines (one encoder $\to$ projection $\to$ BiMamba $\to$
classifier, no cross-attention) reach $0.583$ UA audio-only and $0.652$
text-only --- the bimodal system exceeds the stronger branch by $+7.9$ points
($[-0.094,-0.064]$ for the text-only deficit), so neither modality carries
the system alone. Per-class recall of the featured configuration, computed from the
replication campaign's prediction caches (grand mean $0.721$, not the
$0.730$ headline campaign; three-seed means:
anger $0.798$, happy/excited $0.712$, sadness $0.792$, neutral $0.584$)
locates the residual errors in the neutral class, whose confusions spread
toward happy/excited and sadness --- the classic low-arousal ambiguity of
IEMOCAP.

\paragraph*{SSL front-end arm}
Finally, replacing the trainable raw-convolution front-end with \emph{frozen}
WavLM-Large~\cite{Chen2022WavLM} features under a learnable softmax-weighted
sum (the standard SUPERB layer-probe protocol~\cite{Yang2021SUPERB}) over its
$25$ hidden states lifts the identical downstream architecture to
$0.766\pm0.013$ UA ($+0.036$ $[+0.028,+0.043]$ over the featured arm; $+0.076$
over the same architecture on frozen Wav2Vec2-base, $d_z{=}3.2$) and the
read-out to CCC $0.743/0.701$ --- with \emph{fewer} trainable parameters
(16.0M; the encoder stays frozen). For context, the WavLM paper's own SUPERB
evaluation reports $70.62\%$ ER accuracy on IEMOCAP for
WavLM-Large~\cite{Chen2022WavLM} under SUPERB's lightweight-probe
protocol~\cite{Yang2021SUPERB}; the SUPERB partition and probe differ from
our LOSO protocol, so the two figures contextualize rather than compare.
Consistent with the SSL-era observation
of Wagner~\emph{et al.}~\cite{Wagner2023Dawn}, the front-end representation,
not the temporal backbone or head, is the binding constraint at this scale;
the distributional head transfers to the stronger front-end unchanged. We
keep the raw-conv system as the featured arm --- its front-end trains from
scratch in $\sim$1M parameters, which is the deployment-oriented operating
point this paper targets --- and report the WavLM arm as the
accuracy-oriented configuration.

\begin{table*}[t]
\centering
\caption{Pre-specified controls and extensions (3 seeds $\times$ 5 LOSO
folds, featured recipe unless stated). $\Delta$UA is the seed-averaged paired
difference vs.\ the featured system with its enumerated session-cluster bootstrap
$95\%$ CI; CCC columns give the continuous read-out (head-appropriate:
regression output, marginal expectations, or center of mass). Indented rows
modify only the named component. Table body generated from the released
statistics files (Data and Code Availability).}
\label{tab:controls}
\footnotesize
\setlength{\tabcolsep}{5pt}
\begin{tabular}{lccccc}
\toprule
Arm & UA & $\Delta$UA [95\% CI] & CCC$_V$ & CCC$_A$ & tests \\
\midrule
featured \texttt{mamba\_dual\_head} & $0.730{\pm}0.003$ & --- & $0.661{\pm}0.013$ & $0.661{\pm}0.005$ & reference \\
\;head 2 $\to$ scalar $(v,a)$ regression & $0.723{\pm}0.010$ & $-0.007\;[-0.019,+0.003]$ & $0.687{\pm}0.008$ & $0.678{\pm}0.013$ & any-aux control / CCC anchor \\
\;head 2 $\to$ two 9-way marginals & $0.725{\pm}0.003$ & $-0.005\;[-0.016,+0.006]$ & $0.673{\pm}0.009$ & $0.675{\pm}0.009$ & jointness control \\
\;anisotropic $\sigma_V{=}0.31,\sigma_A{=}0.44$ & $0.728{\pm}0.005$ & $-0.002\;[-0.010,+0.006]$ & $0.668{\pm}0.012$ & $0.665{\pm}0.012$ & rater-std target width \\
\;front-end $\to$ frozen WavLM-Large & $0.766{\pm}0.013$ & $+0.036\;[+0.028,+0.043]$ & $0.743{\pm}0.004$ & $0.701{\pm}0.004$ & SSL front-end arm \\
audio-only branch & $0.583{\pm}0.007$ & $-0.148\;[-0.179,-0.124]$ & --- & --- & modality ablation \\
text-only branch & $0.652{\pm}0.004$ & $-0.079\;[-0.094,-0.064]$ & --- & --- & modality ablation \\
\;rater-vote mixture target (pilot) & $0.728{\pm}0.007$ & --- & $0.654{\pm}0.009$ & $0.609{\pm}0.012$ & supervision pilot \\
\bottomrule
\end{tabular}

\end{table*}

\subsection{Backbone Cost}\label{sec:results-cost}
For completeness we characterize the cost of each backbone \emph{neutrally};
this is not an efficiency claim. Table~\ref{tab:latency} reports forward+backward
latency for the isolated $6$-layer bidirectional temporal encoders ($d{=}256$,
batch $8$, bf16, median of three runs) on a single RTX~5090 across sequence
length $T$.

\begin{table}[t]
\centering
\caption{Backbone forward+backward latency (ms) vs.\ sequence length $T$
(isolated $6$-layer bidirectional encoder, $d{=}256$, batch~$8$, bf16,
median of 3, RTX~5090). Lower is faster.}
\label{tab:latency}
\resizebox{\columnwidth}{!}{%
\begin{tabular}{rcccc}
\toprule
$T$ & Transformer & Mamba-1 & Mamba-2 & Mamba-3 \\
\midrule
$550$  & $\mathbf{5.4}$  & $12.5$ & $28.5$ & $22.4$ \\
$2048$ & $\mathbf{10.7}$ & $16.6$ & $29.1$ & $26.4$ \\
$2750$ & $\mathbf{16.5}$ & $41.0$ & $29.2$ & $36.0$ \\
$4096$ & $\mathbf{29.0}$ & $33.7$ & $30.6$ & $53.5$ \\
$8192$ & $89.1$ & $67.1$ & $\mathbf{63.9}$ & $104.9$ \\
\bottomrule
\end{tabular}}
\end{table}

The measured picture is the opposite of the usual ``Mamba is faster'' narrative at
the lengths this task actually uses. \emph{(i)} At every IEMOCAP-relevant length
($T\!\le\!2750$) the Transformer is the \emph{fastest} backbone, because PyTorch's
fused (flash) attention keeps its constant-factor low; the quadratic cost only
catches up at $T\!\approx\!4096$ and overtakes Mamba only at $T\!=\!8192$.
\emph{(ii)} Peak activation memory tells the same story: with fused attention the
Transformer's peak memory stays the \emph{lowest} of all four backbones even at
$T\!=\!8192$ ($\sim$$4.0$\,GB vs.\ $4.4$/$5.0$/$11.6$\,GB for Mamba-1/2/3), so
there is no ``$\mathcal{O}(T^2)$ memory wall'' to exploit here. \emph{(iii)}
Among the state-space backbones, Mamba-2 is the most predictable --- its latency
is essentially \emph{flat} ($\sim$$29$\,ms) up to $T\!\le\!4096$ because its
chunked-matmul form amortizes a fixed overhead (at $T\!=\!8192$ it rises to
$63.9$\,ms, Table~\ref{tab:latency}), whereas Mamba-1 hits a reproducible slow-kernel
band at the non-power-of-two lengths around $T\!\approx\!2750$. \emph{(iv)}
Mamba-3 is the costliest at long $T$ on both latency and memory. We therefore do
\emph{not} sell a latency advantage; the case for the state-space backbone rests
on its small (not statistically established) accuracy edge
(Sec.~\ref{sec:results-main}) and its linear-memory
recurrence, with any latency benefit deferred to the much longer inputs targeted
in future work.

\section{Discussion and Limitations}\label{sec:discussion}
We state the limitations up front, since several of them shape how the results
should be read.

\paragraph*{The soft target is a constructed prior, not measured uncertainty}
This is the limitation we put first. The Gaussian target of
Eq.~\eqref{eq:soft-target} applies the same researcher-chosen width $\sigma$ to
every utterance, centered on the released (rater-averaged) annotation; it does
not use per-rater ratings, and the model is therefore never supervised with
empirical annotator disagreement. Consequently, claims about the predicted
distribution must stay representational: the head exposes a normalized,
possibly multi-peaked output over the V--A grid, but a diffuse prediction is
not, by itself, strong evidence of a genuinely mixed or ambiguous state. A related
artifact is structural: because the grid is bounded, targets near the edge of
$[1,5]^2$ are truncated and carry less entropy than central ones, so target
entropy correlates with the annotation's distance to the grid edge --- entropy
comparisons across classes must control for this. We tested the link between
predicted entropy and human disagreement directly
(Sec.~\ref{sec:results-analysis}): entropy shows a marginally \emph{negative}
association with \emph{dimensional} per-rater V--A spread (session bootstrap
$[-0.09, -0.00]$; a two-to-three-rater noise floor on
IEMOCAP), but a small, grid-edge-controlled \emph{positive} association with
\emph{categorical} rater ambiguity (partial $r\!\approx\!0.07$--$0.09$, positive
across seeds). The predicted diffuseness thus carries a modest, genuine trace of
categorical annotator ambiguity, though it is far too weak to read as a
calibrated uncertainty estimate. Supervising the head with the per-rater
\emph{category} distribution --- so that the target's spread reflects measured
ambiguity by construction, the perspectivist position of
learning-from-disagreement research~\cite{Davani2022Disagreements} --- rather
than a fixed Gaussian width is the most
direct way to strengthen this link, and is the focus of our continuing work.
Two further design constants deserve the same scrutiny. The width is
\emph{isotropic} ($\sigma_V\!=\!\sigma_A\!=\!0.5$) although our own per-axis
rater-spread estimates are anisotropic ($0.31$ valence vs.\ $0.44$ arousal,
Sec.~\ref{sec:method-vahead}); we ran the rater-matched per-axis variant and
found it changes neither accuracy nor the read-out
(Sec.~\ref{sec:results-controls}) --- the isotropic default is not a
liability here. And the $9$-point resolution,
while unswept, is not arbitrary: the annotations were collected with 5-point
Self-Assessment-Manikin-style scales~\cite{Bradley1994SAM}, and 2--3-rater
averages on a 5-point scale have support on half-integer steps --- exactly the
$9$ points the grid provides per axis. A third commitment is \emph{scale
type}: the grid's numeric coordinates, the metric Gaussian target, and the
center-of-mass and CCC read-outs all treat the SAM-derived rater means as
interval-scaled. The ordinal-affect position~\cite{Yannakakis2021Ordinal}
holds that emotion annotations are more faithfully treated as ranks; under a
monotone re-scaling of the axes the categorical head and UA are unchanged,
whereas the Gaussian width, the expectation read-out, and CCC are not
rank-invariant, so the continuous results should be read within the interval
convention they share with the CCC-based dimensional-SER literature, not as
scale-free findings.

\paragraph*{Scale and SOTA}
All experiments run on a single GPU with frozen large encoders, and the current
results are on IEMOCAP only. We do not claim to beat every reported IEMOCAP
number; the higher figures use \emph{different} evaluation protocols --- random
utterance-level splits that leak speaker identity, or other speaker-independent
partitions (5-fold cross-validation, leave-one-speaker-out; note that
leave-one-\emph{speaker}-out is weaker than session-out on IEMOCAP, since the
held-out speaker's dialogue partner remains in training) --- that are not
directly comparable to our rotating-validation LOSO, and among methods under a
comparably strict speaker-independent protocol our results are competitive
(Sec.~\ref{sec:related}). The frozen-encoder choice is itself a scoping
decision: SSL encoders are the decisive accuracy lever in modern SER
and reach strong valence through implicitly learned linguistic
content~\cite{Wagner2023Dawn} --- a claim our own SSL arm now quantifies
in-pipeline: frozen WavLM-Large features lift the identical architecture by
$+3.6$ UA points and the valence read-out by $+0.08$ CCC
(Sec.~\ref{sec:results-controls}), so the front-end, not the backbone or
head, is the binding constraint at this scale. Our contribution is the
representation and the protocol, not a leaderboard maximum.

\paragraph*{Sensitivity of the V--A head}
The head has two free design parameters, the grid resolution ($9\times9$) and the
soft-target width $\sigma$ (with the mixing weight $\lambda{=}0.3$ fixed). We
ablated $\sigma$ and found it a controllable, near-cost-free knob on
distributional sharpness (Sec.~\ref{sec:results-vahead}); grid resolution we did
not sweep, and a finer grid would trade sharper localization against more
sparsely-supervised cells. The predicted distributions are also \emph{diffuse}
per utterance (entropy $\sim$$77\%$ of maximum at $\sigma{=}0.5$) and only weakly
selective between correct and incorrect predictions; the aggregate geometry is
clean (Fig.~\ref{fig:va-heatmaps}) but per-sample sharpness is modest. Finally,
the classifier is over-confident (ECE~$\approx\!0.20$) --- an orthogonal issue
that the validation-fitted temperature of Sec.~\ref{sec:results-analysis}
already repairs post hoc to ECE~$\approx\!0.02$ without changing any decision.
A natural next step, left to future work, is to promote the V--A matrix from an
auxiliary output to the primary decision --- for instance by self-distilling the
categorical head onto the grid --- so that the distribution itself, rather than a
parallel 4-way head, drives the prediction.

\paragraph*{Cross-corpus harmonization is an assumption, not a result}
The harmonization onto one V--A grid (per-axis affine normalization $+$ dropping
Dominance) is well-defined but unproven across corpora; the cross-corpus
generalization experiment it enables awaits \emph{training} on the MSP corpora
(Sec.~\ref{sec:setup-data}; the zero-shot IEMOCAP$\to$MSP-IMPROV probe below
is a first, deliberately untuned step), and cross-lingual transfer in particular is known
to require dedicated strategies such as SSL layer-anchoring~\cite{Upadhyay2024LayerAnchoring}. Treating Pleasure as Valence and discarding
Dominance is a deliberate lossy projection that may not hold equally for acted
and naturalistic corpora --- and it is lossy in a clinically relevant way:
affective-science evidence indicates the emotion space is not two-dimensional,
with potency/dominance the axis that separates otherwise co-located states such
as anger and fear/anxiety~\cite{Fontaine2007World}. Relatedly, the counseling
motivation of Sec.~\ref{sec:intro} must be read with the inferential limits of
expression-based emotion recognition in mind~\cite{Barrett2019Reconsidered}:
this system estimates the affect that raters would perceive from voice and
words, under an assistive, human-in-the-loop framing --- it does not diagnose
internal states, and our evidence base (acted, English, ten speakers, reference
transcripts) licenses no direct deployment inference.

\paragraph*{State-space efficiency does not materialize here}
We are explicit that at IEMOCAP utterance lengths the state-space backbones are
\emph{not} faster than a flash-attention Transformer, and do not have lower peak
memory (Sec.~\ref{sec:results-cost}). The accuracy edge is real but small. We
therefore frame Mamba as a competitive, fast-converging, linear-memory backbone
whose latency advantage is expected only at the much longer inputs of future
work, not as an efficiency win at the present operating point. Relatedly, the
Mamba-3~\cite{Lahoti2026Mamba3} variant we could train underperforms here ---
with an important scope qualifier: its most expressive component
(MIMO) cannot be \emph{trained} on our consumer Blackwell GPU, since while the
bf16 forward kernel fits, the backward kernel's dynamic shared-memory request
floors at $\sim$$123$\,KB across chunk sizes and ranks, above the
$\sim$$100$\,KB per-block opt-in --- so we evaluate a reduced SISO variant, and
our negative result speaks to that variant in this setting, not to Mamba-3 in
general. Its headline gains target long-context language modeling,
state-tracking, and decode-time efficiency; replacing the short causal
convolution of Mamba-1/2 with RoPE also removes a local mixing that appears
useful for frame-local prosody. None of this helps short, pooled,
speaker-independent SER, so we adopt Mamba-2 as the backbone and report the
SISO Mamba-3 result as a controlled negative.

\paragraph*{Pre-specified revision experiments}
Every analysis and training run pre-specified for this revision has been
executed and is reported in Sec.~\ref{sec:results-controls}
(Table~\ref{tab:controls}): the null-model controls, the rater-vote mixture
pilot, the anisotropic soft target, the frozen-WavLM SSL arm, and the
unimodal baselines with per-class recall --- including the results that
narrow our claims (the aux-form-agnostic classification lift, the regression
head's higher CCC, and the mixture pilot's arousal cost).

\paragraph*{Cross-corpus and demographic probes}
Two further probes scope external validity. \emph{Zero-shot cross-corpus:}
scoring the featured arm unchanged on all $7{,}798$ four-class utterances of
MSP-IMPROV (a $44.1$\,kHz acted corpus whose R/T scenarios repeat fixed target
sentences, making the text branch lexically uninformative there) yields
$45.6\%$ UA (chance $25\%$; seed spread $<\!0.1$ points), with CCC $0.29$ for
valence and $0.35$ for arousal after harmonizing IMPROV's inverted arousal
scale ($1{=}$active $\dots$ $5{=}$calm --- established corpus-internally: on
the raw scale ``angry'' would be rated calmer, $2.76$, than ``neutral,''
$3.86$). Spontaneous scenarios transfer far better than fixed-sentence ones
(UA $0.49$/$0.42$ vs.\ $0.37$/$0.28$), and the weak in-domain
entropy--ambiguity link vanishes out of domain ($\rho\!\approx\!0.02$). We
present this as distribution transfer under a known domain gap, not as a tuned
cross-corpus result. \emph{Demographics:} on the held-out IEMOCAP sessions the
male--female accuracy gap is not robust ($\Delta$UA $+0.010$, session-bootstrap
interval spans zero), but the arousal read-out is markedly stronger for female
speakers (CCC $0.70$ vs.\ $0.62$) --- a monitoring-relevant disparity that any
deployment should track, which we flag rather than explain post hoc.

\paragraph*{Governance and consent}
The intended use --- decision support for a human counselor, never autonomous
intervention --- must also be read against emerging AI governance. The EU AI
Act~\cite{EU2024AIAct} prohibits emotion-recognition systems outright in
workplace and education settings (Art.~5(1)(f); medical and safety uses
excepted) and otherwise treats emotion recognition as a high-risk application
subject to transparency and oversight obligations; a counseling deployment of
this system would sit in that high-risk regime. The scientific-validity
cautions raised against emotion inference in
general~\cite{Barrett2019Reconsidered,Stark2021Ethics} argue for exactly the
posture adopted here: a distributional output that preserves uncertainty and
disagreement rather than asserting a single ground-truth emotion, consumed by
a clinician rather than acted on autonomously. Digital mental-health ethics
additionally requires informed consent, data protection, and clear
accountability lines before any patient-facing
use~\cite{Fiske2019Robot,MartinezMartin2018Apps}, and risk-management
practice such as the NIST AI RMF~\cite{NIST2023AIRMF} supplies the
operational frame (map, measure, manage) for the monitoring obligations this
section has flagged --- the demographic disparity in the arousal read-out
above being a concrete example of what deployment-time measurement must
track.

\paragraph*{Toward Phase 2 (future work)}
This paper is Phase~1: text$+$speech only. The longer-term system we are building
adds a physiological (ECG) stream~\cite{Katsigiannis2018DREAMER} and routes the
$9\times9$ V--A matrix --- rather than a hard class --- as a soft-prompt prefix to a
language model~\cite{Raiaan2024LLMReview} for empathetic response
generation~\cite{Yang2024Empathetic} in counseling support. Bidirectional Mamba has
already shown strong results on ECG/EEG biosignal classification (e.g.\ BioMamba~\cite{Qian2025BioMamba}),
suggesting our BiMamba backbone should transfer to the physiological stream.
Adapting that language model with parameter-efficient methods such as
DoRA~\cite{Liu2024DoRA}, and guaranteeing the deterministic, reproducible
inference~\cite{He2025Nondeterminism} that clinical deployment demands, are
explicit design targets; distribution-free conformal coverage sets over the
grid~\cite{Angelopoulos2023Conformal} are the natural consumable form of
$\mathbf{P}$ for a human-in-the-loop practitioner. Those components are \emph{not} implemented or evaluated
here; the probabilistic, LLM-consumable form of the V--A head is, however,
designed with that downstream interface in mind.

\section{Conclusion}\label{sec:conclusion}
We presented a multimodal (text$+$speech) emotion recognizer that augments the
usual hard classifier with a $9\times9$ probability matrix over the
Valence--Arousal plane, trained with a two-dimensional Gaussian soft target under
a KL$+$cross-entropy objective, and evaluated it under a deliberately strict
speaker-independent LOSO protocol with rotating-session inner validation. Three
findings stand out. First, the probabilistic head is not a tax on accuracy: at
the long operating point its fine-tuned raw-conv configuration reaches
$0.730\pm0.003$ UA over three seeds on IEMOCAP (a separate three-seed
rerun of the same configuration: $0.721\pm0.002$,
Table~\ref{tab:va-loss-ablation}; pooled over both campaigns
$0.726\pm0.006$, $n{=}6$ runs), exceeding the Transformer fusion baseline
significantly under both a paired $t$-test and an enumerated session-cluster
bootstrap (with only five session clusters the strictest cluster-robust
tests cannot reach $p<0.05$; see Sec.~\ref{sec:setup-protocol}), and at
both operating points the dual-head variants are the top configurations --- while
additionally producing an affect distribution whose center of mass tracks
continuous valence and arousal ($r=0.69/0.69$, CCC~$0.66/0.66$;
predominantly between-class structure, within-class $r\!\approx\!0.27$
valence $/$ $0.52$ arousal) and whose per-class means recover
the circumplex geometry (the raw categorical confidence is over-confident,
ECE~$\approx\!0.20$, but a validation-fitted temperature repairs it to
ECE~$\approx\!0.02$ without changing any decision). Second, within one fixed pipeline at matched depth and
width the state-space (Mamba-1/2) backbones are numerically on par with or
slightly above a Transformer; on our hardware Mamba-2's latency is the most
stable across sequence length and overtakes the Transformer only at
$T\!\approx\!8192$, while the trainable (SISO) Mamba-3 variant gives neither an
accuracy nor an efficiency advantage on
this short-utterance task. Third, the soft-target width $\sigma$ is a near-free
knob on distributional sharpness, and a geometry-aware transport loss did not
improve on KL$+$CE. The representation is designed to feed a downstream language
model as a soft prompt; integrating that, together with a physiological stream,
for empathetic response generation in counseling support is left to future work.
Three nearer-term directions follow directly from this study: trained
cross-corpus extension over the harmonized V--A grid
(Sec.~\ref{sec:discussion}); streaming emission of the distribution over long
conversational audio with explicit detection of emotion \emph{shifts} --- a
setting where soft targets provide a natural change signal and long-sequence
backbones should matter most; and input-adaptive soft targets that learn the
Gaussian's width per utterance rather than fixing it globally.

\section*{Acknowledgments}
The authors gratefully acknowledge the sustained support of the AI
Application Laboratory of the Department of Electrical Engineering,
National Changhua University of Education, directed by Prof.~Wen-Ren Yang,
which funded the GPU computing hardware used for all experiments in this
work. This research did not receive any specific grant from funding
agencies in the public, commercial, or not-for-profit sectors.

\section*{CRediT authorship contribution statement}
\textbf{Tingyi Lin:} Conceptualization, Methodology, Software, Validation,
Formal analysis, Investigation, Data curation, Writing -- original draft,
Writing -- review \& editing, Visualization.
\textbf{Wen-Ren Yang:} Supervision, Project administration, Funding
acquisition, Writing -- review \& editing.
\textbf{Kuanwei Chen:} Conceptualization, Writing -- original draft,
Writing -- review \& editing.

\section*{Declaration of competing interest}
The authors declare that they have no known competing financial interests or
personal relationships that could have appeared to influence the work
reported in this paper.

\section*{Data and code availability}
The complete training, evaluation, and analysis code --- including every
table and figure generator and the dataset-index regeneration scripts ---
is publicly available at
\url{https://github.com/brian10420/EchoMind-Mamba-VA-SER} and is archived on Zenodo
(DOI: \href{https://doi.org/10.5281/zenodo.21904810}{10.5281/zenodo.21904810}). The IEMOCAP corpus itself is available from the
University of Southern California under its release license and cannot be
redistributed by us; the repository regenerates our exact utterance index
and splits from a licensed copy, so all reported results are reproducible
end-to-end.

\section*{Declaration of generative AI and AI-assisted technologies in the
manuscript preparation process}
During the preparation of this work the authors used Claude (Anthropic),
accessed through the Claude Code environment across its successive
2025--2026 model versions, in order to assist with LaTeX drafting and
editing of the manuscript. Beyond the writing process, the same assistant
was used as a programming aid in the development of the training,
statistical-analysis, and figure/table generation code
(Sec.~\ref{sec:setup-protocol}); all of this code is released for
inspection (see Data and code availability), and its correctness was
verified by the authors through the released regression-test suite, the
automated gates that regenerate every table cell from the committed
per-run statistics, and independent re-runs of the reported experiments.
After using this tool, the authors reviewed, tested, and edited the
content as needed --- all experiments, analyses, and scientific claims
were verified by the authors --- and the authors take full responsibility
for the content of the published article.

\bibliographystyle{IEEEtran}
\bibliography{refs}

\end{document}